\documentclass[aps,prd,twocolumn,floats,nofootinbib,longbibliography,showkeys]{revtex4-1}
\usepackage[dvips]{graphicx}

\usepackage{amsmath}
\usepackage{braket}
\usepackage{amsfonts}
\usepackage[colorlinks=true]{hyperref}
\usepackage{hyperref}
\usepackage{xcolor}

\usepackage{booktabs}

\usepackage{aas_macros}

\usepackage{caption}
\usepackage{graphicx}
\usepackage{subcaption}

\usepackage{bm}
\usepackage{slashed}

\usepackage{changes}

\def\aap{\textit{Astronomy and Astrophysics}}

\def\be{\begin{equation}}
\def\ee{\end{equation}}
\def\bea{\begin{eqnarray}}
\def\eea{\end{eqnarray}}

\def\Lm{\mathcal{L}_m}

\def\t{\tilde}

\newcommand\rs{\mathit{r}_s}

\definechangesauthor[name={Olivier}, color=magenta]{OM}
\definechangesauthor[name={Aurelien}, color=cyan]{AH}

\begin{document}
\title{Compact objects with scalar charge embedded in a magnetic or electric field in Einstein-Maxwell-dilaton theories}

\author{Olivier Minazzoli}
\email[]{ominazzoli@gmail.com}
\affiliation{Artemis, Universit\'e C\^ote d'Azur, CNRS, Observatoire C\^ote d'Azur, BP4229, 06304, Nice Cedex 4, France,\\Bureau des Affaires Spatiales, 2 rue du Gabian, 98000  Monaco.}
\author{Maxime Wavasseur}
\affiliation{Departament de Física Quántica i Astrofísica (FQA), Universitat de Barcelona (UB), Carrer de Martí i Franquès, 1, Barcelona, 08028, Spain,\\Artemis, Universit\'e C\^ote d'Azur, CNRS, Observatoire C\^ote d'Azur, BP4229, 06304, Nice Cedex 4, France.}
\begin{abstract}
In this paper, we generalize the Schwarzschild-Melvin solution within Einstein-Maxwell-dilaton theories to include non-null scalar charges, while remaining embedded in a magnetic or electric field \textit{à la Melvin}. We then use this general solution to obtain the solution in the specific case of Entangled Relativity after a conformal transformation. This notably enables us to verify that the analytical solution used in [Arruga \& Minazzoli 2021] in order to represent compact objects such as neutron stars in Entangled Relativity is indeed a good approximation of the exact solutions of Entangled Relativity when the background field goes to zero.
\end{abstract}
\maketitle

\section{Introduction}

The phenomenology of Einstein-Maxwell-dilaton theories has been studied a lot following the first superstring revolution, especially with respect to their black-hole solutions \cite{gibbons:1988nb,garfinkle:1991pr,holzhey:1992nb}, sometimes even with the hope of explaining elementary particles as actually being elementary black-holes within this framework \cite{holzhey:1992nb}. Indeed, the low energy effective action of bosonic string and then superstring theories always seemed to imply the presence of the massless scalar particle named \textit{dilaton}, and compactification from either 26 or 10 dimensions to the observed 4 also implies the existence of a plethora of scalar fields, named \textit{moduli} \cite{douglas:2007rm,tong:2009ar}. While arguments has been put forth that such scalar fields acquire a large mass in the process of the \textit{stabilization of the moduli} \cite{vanriet:2024pr,douglas:2007rm}, it may not necessarily be so, although it is nowadays usually considered that they should at least be light instead of massless anyway \cite{douglas:2007rm,burgess:2020fp}.\footnote{Let us note that whether moduli are light or massless can completely change their phenomenology, irrespectively of their ``lightness''. See, for instance, \cite{hees:2018pr}.} Similarly, Kaluza-Klein theories also always imply at least one additional massless scalar-field \cite{vanriet:2024pr}. 

However, in the present paper, we are not so much interested by this historical context, but more by the fact that Entangled Relativity
can be re-written as an Einstein-Maxwell-dilaton theory after a conformal transformation \cite{minazzoli:2021ej,arruga:2021ep}.
Entangled Relativity is an $f(R,\Lm)$ theory, similar to General Relativity, but it adopts a non-linear coupling function, $f(R,\Lm) = \Lm^2/R$, instead of the usual linear Einstein-Hilbert coupling function, $f(R,\Lm) = R/(2\kappa) + \Lm$. The specific form of this function is determined by the requirement that the theory reproduces the phenomenology of General Relativity in the weak-density regime and whenever $\Lm = T$ on-shell. These requirements ensure consistency with the well-tested phenomenology of General Relativity \cite{wambsganss:1998lr,uzan:2011lr,will:2014lr,burns:2020lr,abbott:2020lr,fienga:2024lr}, while allowing for significant deviations at much higher density and pressure---such as those encountered inside a collapsing object forming a black hole or in the early universe---where General Relativity is argued to be insufficiently accurate \cite{penrose:1965pl,senovilla:2012bk,senovilla:2022pt,joshi:2014bk}.
A series of studies have demonstrated that Entangled Relativity indeed reproduces the phenomenology of General Relativity for energy densities lower than those of neutron stars \citep{ludwig:2015pl,minazzoli:2018pr,arruga:2021pr,arruga:2021ep,minazzoli:2021ej,minazzoli:2021cq,wavasseur:2025gg,minazzoli:2025ep,chehab:2025hl}, while predicting small but perhaps eventually measurable deviations for neutron stars, which are notably systematically more massive than in General Relativity, or for white dwarfs \cite{arruga:2021pr,arruga:2021ep,chehab:2025hl}.
The non-linear nature of the coupling function implies that Entangled Relativity relies on fewer dimensionful fundamental constants than General Relativity. Notably, neither $G$ nor $\hbar$ appear in the path-integral formulation of the theory, meaning that neither is strictly constant. Instead, both are related to an additional scalar degree of freedom arising from the non-linearity of $f$ \cite{ludwig:2015pl,minazzoli:2022ar,chehab:2025hl}. Consequently, Entangled Relativity is not only an accurate description of many observed phenomena and a potential improvement in domains where General Relativity is challenged, but it is also a more economical theory in terms of fundamental constants and natural units \cite{minazzoli:2022ar,chehab:2025hl}.\footnote{It is worth emphasizing that this also means Entangled Relativity has no free parameters that can be fine-tuned to match observations, unlike most alternative gravitational theories \cite{will:2014lr,fienga:2024lr}.} As a result, one could argue that Occam's razor favors it.
The issue of cosmic acceleration remains unresolved within the framework of Entangled Relativity \cite{minazzoli:2018pr,minazzoli:2021cq,minazzoli:2024mo}. However, this might simply reflect the comparatively recent development of the theory, which has not yet received the same level of scrutiny or attention as General Relativity. Interestingly, recent developments suggest that this question is also far from settled in the context of General Relativity with a cosmological constant, especially in light of the persistent Hubble tension observed across various independent probes \cite{verde:2024an,desi:2025ar}, as well as other emerging cosmological anomalies \cite{peebles:2025pt}. Notably, whereas $\Lambda$CDM is a mathematical solution of General Relativity, it is incompatible with Entangled Relativity \cite{minazzoli:2018pr,minazzoli:2021cq,minazzoli:2024mo}. As such, mounting evidence against $\Lambda$CDM would place Entangled Relativity on a more comparable footing with General Relativity in addressing this cosmological issue. 
Moreover, as if this were not already a strong motivation to investigate the theory, Entangled Relativity also better aligns with Einstein’s \textit{principle of relativity of inertia} \citep{einstein:1917co,einstein:1918an,einstein:1918sp,einstein:1921bk,pais:1982bk,hoefer:1995cf,minazzoli:2024pp}, which Einstein referred to as \textit{Mach’s principle} in \citep{einstein:1918an}. This principle states that a truly satisfactory theory of relativity should not permit the existence of vacuum spacetimes \citep{einstein:1917co,einstein:1918an,einstein:1918sp,einstein:1921bk,pais:1982bk,hoefer:1995cf}. Indeed, allowing for vacuum solutions would imply that inertia---determined by the metric tensor in relativistic theories---could exist even in the total absence of matter. However, such a scenario would, in essence, contradict the \textit{principle of relativity of inertia} \citep{einstein:1917co,einstein:1918an,einstein:1918sp,einstein:1921bk,pais:1982bk,hoefer:1995cf,minazzoli:2024pp}. The non-linearity of the coupling function in Entangled Relativity $f(R,\Lm) = \Lm^2/R$ prevents even the definition of the theory without matter, thereby aligning more closely with this principle than General Relativity \cite{einstein:1918sp,pais:1982bk,hoefer:1995cf,minazzoli:2024pp}. The theory was named after this property \cite{arruga:2021pr}, as curvature and matter are, in the etymological sense, \textit{entangled} at the level of the theory's formulation.

Given the historical context, many solutions exist already for Einstein-Maxwell-dilaton theories, e.g. \cite{gibbons:1988nb,garfinkle:1991pr,dowker:1994pr,junior:2022pr}. However, to our surprise, we could not find solutions of objects with scalar charges embedded in a magnetic or an electric field, ``à la Melvin'', in Einstein-Maxwell-dilaton theories---although they have been found in General Relativity with a minimally coupled scalar-field \cite{astorino2013pr}. \footnote{Originally, the Schwarzschild-Melvin spacetime is an exact solution of the Einstein electrovacuum equations that describes a black hole immersed in a magnetic field asymptotically aligned with the z-axis \cite{griffiths:2009bk}. This solution is significant for understanding the interaction between geometry and matter and is frequently employed as a model for astrophysical environments \cite{cardoso:2024ar}.} Even more, it has be claimed that the dilaton Melvin-Schwarzschild was 
the only static asymptotically dilaton Melvin axisymmetric black-hole solution in Einstein-Maxwell-dilaton theories \cite{rogatko:2016pr}. Hence, it seemed that such solutions may not exist.

The present paper gives such solutions. However, this is not in contradiction with the proof given in \cite{rogatko:2016pr} per se, because such solutions either represent the exterior solution of \textit{matter-filled} compact objects such as neutron stars or white dwarfs, embedded in an external electric or magnetic field, or they represent naked singularities---which should not be able to form during the collapse of a star into a black-hole because the scalar-charge is radiated away during the collapse via the emission of monopolar gravitational waves \cite{scheel:1995pr,gerosa:2016cq}.

As mentioned above, among the several theories corresponding to an Einstein-Maxwell-dilaton theory, Entangled Relativity is particular, given that it imposes that matter exists in the action, as well as it prevents the existence of vacuum solutions from the onset, because vacuum solutions would be ill-defined. Indeed, the field equations are such that there exists an additional scalar degree-of-freedom that corresponds to the ratio between $R$ and $\Lm$, which is obviously ill-defined when $\Lm=R=0$. 
However, it has been found already in several examples that the ratio remains well-defined when one considers the existence of matter fields from the start and then make their value goes to zero.

Hence, in Entangled Relativity, one must always consider a solution in which there is at least one type of matter fields in the entire spacetime: vacuum solutions cannot be exact, but can be good approximations when the density of matter fields goes to zero.\footnote{As far as we know, there does not exist a place in the universe where all the fields would vanish entirely. Consequently, in our opinion, the vacuum assumption (in the classical sense) is not physical from the start, although it may be a good approximation in some situations---just as the Minkowski assumption for spacetimes.} This was notably the argument in \cite{arruga:2021ep} in order to use the hairy spherical solution of scalar-tensor theories---sometimes named the Just solution \cite{damour:1992cq}, or the Fisher-Janis-Robinson-Winicour solution \cite{astorino2013pr}---in order to model a compact object, such as neutron stars or white dwarfs, in Entangled Relativity. 


\section{Field equations and solutions}

In convenient units, the action of Einstein-Maxwell-dilaton theories can be written as follows \cite{garfinkle:1991pr,holzhey:1992nb}:
\bea \label{eq:actionclER}
S=\int d^{4} x \sqrt{-\t g}\left[\t R-2\t g^{\mu \nu}\partial_\mu \varphi \partial_\nu \varphi -  e^{- 2\alpha\varphi}  \t F^2 \right],
\eea
where $\alpha$ takes a specific value for each specific theory. For instance, it takes the value $\alpha=1$ for bosonic strings \cite{holzhey:1992nb}, $\alpha = \sqrt{3}$ for a 5D Kaluza-Klein theory with one dimension compactified on a circle \cite{holzhey:1992nb} and more generally $\alpha = \sqrt{(2+N)/N}$ for a (4$+N$)D Kaluza-Klein theory with a compactified N-dimensional torus \cite{gibbons:1988nb}, whereas $\alpha=1/(2\sqrt{3})$ for Entangled Relativity \cite{minazzoli:2021ej}.

In order to find the new solution of a hairy compact object embedded in a magnetic or an electric field, we start from the already well-known solution of a hairy compact object in vacuum \cite{arruga:2021ep}, and then we apply the procedure of \cite{dowker:1994pr}  that enables to generate new solutions with a magnetic field from axisymmetric solutions. Then we use the invariance of the field equations of Einstein-Maxwell-dilaton theories under a well-known duality rotation \cite{garfinkle:1991pr,holzhey:1992nb} to deduce the corresponding electric case.

The metric solution we obtained for both the electric and magnetic case reads as follows
\bea \label{eq:solmag}
&&d\t s^2 = \left(1 - \frac{\rs}{r}\right)^{1-b} \left[   \frac{r^2 sin^2 \theta}{\Lambda^{\frac{2}{1+\alpha^2}}} d\psi^2 +  \Lambda^{\frac{2}{1+\alpha^2}}r^2 d\theta^2\right]\nonumber \\
&&+\Lambda^{\frac{2}{1+\alpha^2}} \left[\left(1 - \frac{\rs}{r}\right)^{-b}dr^2  - \left(1 - \frac{\rs}{r}\right)^{b} dt^2 \right], 
\eea
where $\rs$ is the Schwarzschild radius and $b \in~ ]0,1]$ is an arbitrary parameter that quantify the amount of scalar charge of the solution.

One can see that this line element corresponds to the exterior solution of a \textit{matter-filled} spherical object of radius $R_*>r_s$, such as a star; or of a naked singularity, because the Ricci scalar diverges at $r=r_s$.\footnote{It is straightforward to check that $\tilde R$ diverges at $r=r_s$ because $\Lambda =1$ in that limit such that one simply recovers the naked singularity of the Just-Fisher-Janis-Robinson-Winicour solution provided in Appendix \ref{sec:FJRW}.}
\subsection{Solution in a magnetic field}
\label{sec:solmag}

In the magnetic case, we found the following solution for the fields other than the metric. The scalar-field reads
\be \label{eq:defvarphi}
\varphi =  \frac{\sqrt{1-b^2}}{2} \ln \left(1 - \frac{\rs}{r}\right) - \frac{\alpha}{1+\alpha^2} \ln \Lambda,
\ee
while the solution for the electromagnetic field is 
\be
A = \frac{2}{1+\alpha^2} \frac{1}{B}\left(1 - \frac{1}{\Lambda}\right) d\psi,
\ee
with
\be
\Lambda := 1 + \frac{1+\alpha^2}{4} \left(1 - \frac{\rs}{r}\right)^{1-b + \alpha \sqrt{1-b^2}} B^{2} r^2 \sin^2\theta,
\ee
corresponding to a magnetic field pointing along the z-direction that reads
\be \label{eq:magfieldEiF}
\mathfrak{B} = \frac{2 \, \frac{\partial\,\Lambda}{\partial {\theta}}}{{B\left(1+a^{2}\right)} \Lambda^{2}}dr - \frac{2 \, \frac{\partial\,\Lambda}{\partial r}}{{B \left(1+a^{2}\right)} \Lambda^{2}}d\theta.
\ee

Let us note that this solution is a generalization of the solution given in \cite{junior:2022pr} when the compact object possesses a scalar charge. While these black-holes with a scalar charge---actually corresponding to naked singularities as we discussed above---are not expected to be able to form from the collapse of a star because they loose their scalar charge through the emission of monopolar gravitational waves during the process \cite{scheel:1995pr,gerosa:2016cq}, this new solution can nevertheless depict the exterior metric generated by a spherical star that would be embedded in an external magnetic or electric field. One can verify that in the $B\rightarrow 0$ limit, we have $A \rightarrow 0$ and the line element converge to the Just-Fisher-Janis-Robinson-Winicour line element.

We verified this solution with SageManifolds \cite{gourgoulhon:2015jc}. The notebook is freely available at
\cite{notebook_EINSTEIN_mag}.

\subsection{Solution in an electric field}
\label{sec:solelec}

The scalar and electromagnetic field solutions in the case of a uniform electric field can be obtained through the following transformation \cite{garfinkle:1991pr,holzhey:1992nb}

\begin{subequations}\label{eq:transfoEF}
\bea
&& F_{\mu \nu}  \longrightarrow F^{e}_{\mu \nu} = \frac{1}{2} e^{-2\alpha \varphi}\epsilon_{\mu \nu \kappa \lambda} F^{\kappa \lambda},\\
&& \varphi \longrightarrow \varphi^{e} = - \varphi,
\eea
\end{subequations}
where $\epsilon_{\mu \nu \kappa \lambda}$ is the Levi-Cività tensor. In this case, the electromagnetic field explicitly reads as follows
\be \label{eq:AelecEF}
A = B\rs \cos{\theta} \left[\left(\frac{1+b}{2}-\frac{\alpha}{2}\sqrt{1-b^2}- \frac{r}{\rs}\right)\right]\mathrm{d} t,
\ee
corresponding to an electric field pointing along the z-direction that reads

\begin{eqnarray} \label{eq: EelecEF}
\mathfrak{E} &=& -\frac{1}{2}\left(2r+\left(a \sqrt{1-b^{2}} - b-1\right) r_{s}\right) B\sin \theta d\theta\nonumber \\
&& + B\cos\theta dr.
\end{eqnarray}

Again, one can verify that in the $B\rightarrow 0$ limit, we have $A \rightarrow 0$ and the line element converge to the Just-Fisher-Janis-Robinson-Winicour line element.

The notebook that verifies this solution is freely accessible at: \cite{notebook_EINSTEIN_elec}.

\subsection{Discussion on the previously claimed uniqueness of the dilaton Melvin-Schwarzschild black-holes}
\label{sec:uniqueness}

It has been claimed not so long ago \cite{rogatko:2016pr} that the dilaton Melvin-Schwarzschild solution---which corresponds to Eq. (\ref{eq:solmag}) for
$b=1$ \cite{yazadjiev:2006pr,junior:2022pr}---should be 
the only asymptotically dilaton Melvin static axisymmetric black-hole solution of Einstein-Maxwell-dilaton theories. While the solutions provided by Eqs. (\ref{eq:solmag}-\ref{eq:AelecEF}) for $b\neq 1$ seem to be counter-examples to that assertion, they are not, because they correspond to naked singularities or \textit{matter-filled} spherical compact objects instead of black-holes. 
\section{Entangled Relativity}
\label{sec:ER}

In convenient units, the action of Entangled Relativity in the frame of its formulation---which we shall hereafter name the ``\textit{entangled frame}''---reads as follows \cite{ludwig:2015pl}
\be \label{eq:SER}
S = -\frac{1}{2}\int d^4 x \sqrt{-g} \frac{\Lm^2}{R}.
\ee
The field equations that follows from the extremization of the action are \cite{ludwig:2015pl}
\be \label{eq:metric}
G_{\mu \nu} = \frac{T_{\mu \nu}}{\vartheta} + \vartheta^{-2} \left[\nabla_\mu \nabla_\nu - g_{\mu \nu} \Box \right] \vartheta^{2},
\ee
with $\vartheta$ a scalar defined upon the ratio between the scalars $\Lm$ and $R$
\bea
\vartheta := - \frac{\Lm}{R},\label{eq:vartheta}
\eea
\footnote{$\vartheta$ has the dimension of $c^4/G$ but can easily be rendered dimensionless by using its expectation value $\vartheta_0 = c^4/(8\pi G)$.} with the following stress-energy tensor
\be
T_{\mu \nu} := -\frac{2}{\sqrt{-g}} \frac{\delta\left(\sqrt{-g} \mathcal{L}_{m}\right)}{\delta g^{\mu \nu}},
\ee
which is not classically conserved since
\be 
\label{eq:noconsfR}
\nabla_{\sigma}\left(\vartheta T^{\alpha \sigma}\right)=\mathcal{L}_{m} \nabla^{\alpha} \vartheta . 
\ee
For an electromagnetic field, the matter field equation on the other hand reduces to
\be
\nabla_\sigma \left(\vartheta F^{\mu \sigma} \right)=0.
\ee

It may na\"ively be thought that Eq. (\ref{eq:vartheta}) becomes singular as $R \rightarrow 0$, but it is not the case, as the value of this ratio must satisfy all the field equations. In particular, one can see that it behaves as a scalar-degree of freedom---as usual in $f(R)$ or $f(R,\Lm)$ theories \cite{capozziello:2015sc,harko:2010ep}---with the following field equation \cite{ludwig:2015pl}
\be
3 \vartheta^{-2} \Box \vartheta^2 = \vartheta^{-1} \left(T - \Lm \right).
\ee
Because this field equation is perfectly well-behaved, the ratio $\Lm/R$ must be perfectly well defined in the limit $R\rightarrow 0$, which should therefore simply correspond to the limit $\Lm \rightarrow 0$, with the same rate of convergence toward zero. This has been verified already for spherically charged black-holes \cite{minazzoli:2021ej}, slowly rotating charged black-holes \cite{wavasseur:2025gg}, neutral spherical black-holes embedded in an electric or a magnetic field \cite{minazzoli:2024ar}. We will verify this for hairy compact objects embedded in an electric or a magnetic field in this paper, as well as provide a general explanation of why it is the case in Sec. \ref{sec:discEF}.

As explained in \cite{minazzoli:2021ej}, in order to deduce the solution for Entangled Relativity defined in Eq. (\ref{eq:SER}) from Einstein-Maxwell-dilaton theories defined in Eq. (\ref{eq:actionclER}), one simply has to use the following conformal transformation 
\be \label{eq:conftransER}
g_{\alpha \beta}=e^{4\alpha \varphi} \t g_{\alpha \beta}\textrm{ with }\alpha = 1/(2\sqrt{3}),
\ee
and $\varphi$ defined in Eq. (\ref{eq:defvarphi}).
\subsection{Solution in a magnetic field}
\label{sec:solmagER}

 In particular, in the magnetic field case, one has
\be
g_{\alpha \beta}=\frac{\left(1-\frac{r_s}{r} \right)^{2\alpha \sqrt{1-b^2}}}{\Lambda^{\frac{4\alpha^2}{1+\alpha^2}}} \t g_{\alpha \beta},
\ee
with
\be
\alpha = \frac{1}{2\sqrt{3}}.
\ee
The solution of the theory defined in Eq. (\ref{eq:SER}) therefore reads as follows
\begin{widetext}
\be\label{eq:msolmag}
ds^2 =  -{\left(1-\frac{r_{s}}{r}\right)}^{b} \Delta^{2} \Lambda^{\frac{20}{13}} dt^2 + \frac{\Delta^{2} \Lambda^{\frac{20}{13}}}{{\left(1-\frac{r_{s}}{r} \right)}^{b}} dr^2 + \left( {\left(1-\frac{r_{s}}{r} \right)}^{1-b} \Delta^{2} \Lambda^{\frac{20}{13}} \right) r^{2}  d\theta^2 + \left( \frac{ {\left(1-\frac{r_{s}}{r} \right)}^{1-b} \Delta^{2} }{\Lambda^{\frac{28}{13}}} \right) r^{2} \sin^{2}{\theta}d\psi^2,
\ee
\end{widetext}
with
\be \label{eq:delta}
\Delta :=  {\left(1-\frac{r_{s}}{r}\right)}^{\frac{\sqrt{1-b^{2}}}{2\sqrt{3}} },
\ee
and where 
\be
\Lambda = 1+ \frac{13}{48} \, B^{2} {\left(1-\frac{r_{s}}{r}\right)}^{\frac{\sqrt{1-b^{2} }}{2\sqrt{3}}+ 1-b}  r^{2} \sin^2{\theta}.
\label{eq:lambdamag}
\ee
The electromagnetic field is
\be \label{eq:Amag}
A = \frac{24}{13 B} \left( 1 - \frac{1}{\Lambda} \right) \mathrm{d} {\varphi},
\ee
corresponding to a magnetic field pointing along the z-direction that reads

\be
\mathfrak{B} = \frac{24 \, \frac{\partial\,\Lambda}{\partial {\theta}}}{13 \, B\Lambda^{2}}dr -\frac{24 \, \frac{\partial\,\Lambda}{\partial r}}{13 \, B \Lambda^{2}}d\theta.
\ee

Using that $\vartheta = e^{-2\alpha \varphi}$ \cite{minazzoli:2021ej}, one also deduces that
\be \label{eq:dofmag}
\vartheta = - \frac{\Lm}{R}= \frac{\Lambda^{\frac{2}{13}}}{\Delta} .
\ee
Hence, one can verify that the ratio between $\Lm$ and $R$ remains perfectly well-defined in the vacuum limit; which corresponds to $B \rightarrow 0$. Indeed, one can check that one has
\be
\Lm = -\frac{B^{2} \Xi\Lambda^{-44/13}}{48r^{2}\Delta^{2}\left(1-\frac{r_{s}}{r}\right)},
\ee
and 
\be
\label{eq: Ricci_scalar_mag}
R = \frac{B^{2} \Xi\Lambda^{-46/13}}{48r^{2}\Delta\left(1-\frac{r_{s}}{r}\right)},
\ee
where 
\bea
\label{eq:Xidef}
\Xi(r,\theta) &=&  48r^2\left(1-\frac{\rs}{r}\right)\\
&-&4\sqrt{3}r r_s  \left( \left( b + 1\right) \frac{r_{s}}{r} - 2 \right) \sqrt{1-b^2} \sin^2\theta \nonumber\\
&-& r r_s\left( 48 b - \left( 11 b^2 + 24 b + 13 \right) \frac{r_{s}}{r} \right) \sin^2\theta. \nonumber
\eea
The notebook that verifies this solution is freely accessible at: \cite{notebook_ER_mag}.

We see that $R$ diverges at $r=r_s$ for $b\in ]0,1[$. Indeed, in Eq.(\ref{eq: Ricci_scalar_mag}), one has $\lim_{r=r_{s}}\Delta=0$, $\lim_{r=r_{s}}\Lambda=1$ and the function $\Xi$ is nonzero and finite for $b\neq1$. Therefore, the solution either corresponds to the exterior solution of a \textit{matter-filled} spherical object with scalar hair, or to a naked singularity, but not to a black-hole.

It is also worthwhile to note that the solution is such that $\Lm<0$ and $R>0$, such that $R/\Lm<0$.
\subsection{Solution in an electric field}
\label{sec:solelecER}

The electric version of the solution can be obtained through the following transformation \cite{minazzoli:2025ep}
\begin{subequations}\label{eq:transfo}
\bea
&&F_{\mu \nu}  \longrightarrow F^{e}_{\mu \nu} =-\frac{1}{2} \frac{\Lm}{R} ~ \epsilon_{\mu \nu \kappa \lambda} F^{\kappa \lambda},\\
&&g_{\mu \nu} \longrightarrow g^e_{\mu \nu} = \left(\frac{\Lm}{R}\right)^4 g_{\mu \nu},
\eea
\end{subequations}\label{eq:litdiffm}
where $\epsilon_{\mu \nu \kappa \lambda}$ is the Levi-Cività tensor, such that \cite{wavasseur:2025gg}
\be \label{eq:transfoERmag2elec}
\vartheta \longrightarrow \vartheta^e = \frac{1}{\vartheta}.
\ee
Let us stress that unlike in the Einstein frame, the line element in the electric case differ from the magnetic one.
 Hence, in the case of a uniform electric field, the solution for the metric is

\begin{widetext}
\be\label{eq:msolelec}
ds^2 =  -{\left(1-\frac{r_{s}}{r}\right)}^{b}  \frac{\Lambda^{\frac{28}{13}}}{\Delta^{2}} dt^2 + \frac{ \Lambda^{\frac{28}{13}}}{\Delta^{2}{\left(1-\frac{r_{s}}{r} \right)}^{b}} dr^2 + \left( {\left(1-\frac{r_{s}}{r} \right)}^{1-b} \frac{\Lambda^{\frac{28}{13}}}{\Delta^{2}} \right) r^{2}  d\theta^2 + \left( \frac{ {\left(1-\frac{r_{s}}{r} \right)}^{1-b} }{ \Delta^{2} \Lambda^{\frac{20}{13}}} \right) r^{2} \sin^{2}\theta d\psi^2,
\ee
\end{widetext}
where the quantities $\Lambda$ and $\Delta$ are given in Sec. \ref{sec:solmagER}.

The solution for the electromagnetic 4-vector is
\be \label{eq:Aelec}
A = B\rs \cos{\theta} \left[\left(\frac{1+b}{2}-\frac{1}{4\sqrt{3}}\sqrt{1-b^2}- \frac{r}{\rs}\right)\right]\mathrm{d} t,
\ee
corresponding to an electric pointing along the z-direction that reads

\begin{eqnarray}
\mathfrak{E} &=& \left(r-\frac{1}{2} r_{s}\left(b+1-\frac{\sqrt{1-b^{2}}}{2\sqrt{3}}\right) \right)B\sin {\theta} d\theta \nonumber\\
&& -B\cos {\theta}dr
\end{eqnarray}

We obtain in this case
\be \label{eq:dofelec}
\vartheta = - \frac{\Lm}{R}= \frac{\Delta}{\Lambda^{\frac{2}{13}}} .
\ee
Which is the inverse of the magnetic scalar-field solution given in Eq. (\ref{eq:dofmag}) ---as it could have been anticipated from Eq. (\ref{eq:transfoERmag2elec}).
We can verify in this case again that the ratio between $\Lm$ and $R$ remains perfectly well-defined in the vacuum limit---as it can be deduced from their respective values:
\bea
&&\Lm = \frac{B^{2} \Xi \Delta^{4}}{48\left(1-\frac{\rs}{r}\right)r^2\Lambda^{56/13}},\\
&&R = -\frac{B^{2} \Xi \Delta^{3}}{48\left(1-\frac{\rs}{r}\right)r^2\Lambda^{54/13}},
\eea
where $\Xi$ is defined in Eq. (\ref{eq:Xidef}). The notebook that verifies this solution is freely accessible at: \cite{notebook_ER_elec}.

As in the magnetic case, one also has a naked singularity at $r=r_s$. Indeed, $\lim_{r=r_s}\Delta^{3}/(1-r_s/r)$ diverges because ${\frac{\sqrt{3}}{2}\sqrt{1-b^{2}}-1}$ is negative for $b \in ]0,1[$, while $\lim_{r=r_{s}}\Lambda=1$ and the function $\Xi$ is nonzero and finite for $b\neq1$, such that $R$ diverges at $r=r_s$ for $b\neq 1$.

Since the magnetic and electric metric solutions differ only by a conformal factor, both cases should yield a similar Petrov classification. We therefore verify that the spacetime is algebraically general in the external region in the case of an electric field (see \cite{notebook_ER_petrov}).

It is also worthwhile to notice that the solution is such that $\Lm>0$ and $R<0$---that is, the opposite situation with respect to the magnetic case---such that one still has $R/\Lm<0$ nonetheless.
\subsection{Discussion}
\label{sec:discEF}

 It is important to note that, unlike in the Einstein-frame where the solution for $F^2=0$ induces $\t R \neq 0$ because of the dilaton field, the solution in the Entangled frame is such that $R \rightarrow 0$ for $\Lm \rightarrow 0$---see Appendix \ref{sec:FJRW}. This can be understood from the following consideration. The Lagrangian density can be rewritten as follows
\be \label{eq:SER2}
-\frac{1}{2} \frac{\Lm^2}{R}= \frac{\Lm^2}{R^2} \frac{R}{2} - \frac{\Lm}{R} \Lm = \vartheta^2 \frac{R}{2}+\vartheta \Lm.
\ee
Also, as a matter of fact, $\vartheta$ can be considered as an independent scalar field in the right-hand-side of Eq. (\ref{eq:SER2}) when considered as a standalone Lagrangian density, because the resulting field equations are completely equivalent to the field equations derived from the left-hand-side of Eq. (\ref{eq:SER2}) \cite{ludwig:2015pl}. Hence, for any non-null $\vartheta$ field, since the action is minimal for its solutions, one must have $R \rightarrow 0$ for $\Lm \rightarrow 0$ on-shell, just as in General Relativity. This shows that all the vacuum solutions of General Relativity are limits of solutions of Entangled Relativity for vanishing matter fields---but not necessarily the only limits however. Moreover, given that $\vartheta$ can take any finite initial value for all such solutions, it necessarily means that $R$ and $\Lm$ converges to zero at the same rate while reaching such limiting solutions in Entangled Relativity. But of course, one would not be allowed to start from a solution that is such that $\Lm=0=R$ from the beginning in Entangled Relativity, as $\vartheta = - \Lm/R$ would be ill-defined in that situation.\footnote{Let us emphasize that, in our opinion, having $\Lm=0=R$ is pure fantasy anyway because, as far as we know, there does not exist a single location in our universe where it would be \textit{exactly} true.} Therefore, we still argue that Entangled Relativity better reflects Einstein's ``Mach's Principle''---which states that there can be \textit{no spacetime without matter} \cite{einstein:1918an,einstein:1918sp,hoefer:1995cf,minazzoli:2024pp}---than General Relativity.

Also, it is important to stress that the vacuum solutions of General Relativity are not the only limits for vanishing matter fields, as can be probed explicitly with the solution provided in Eqs. (\ref{eq:msolmag}-\ref{eq:Amag}) and  Eqs. (\ref{eq:transfoERmag2elec}-\ref{eq:Aelec})---because the limit $A \rightarrow 0$ is not necessarily the Schwarzschild metric, but can be an object with scalar hair. In particular, both solutions have the metric considered in \cite{arruga:2021ep} [their Eqs. (22-26)] to describe the exterior solutions of compact objects with scalar hair in Entangled Relativity as limit when $A\rightarrow 0$. Therefore, as presented in the abstract, it shows that the analytical solution used in \cite{arruga:2021ep} in order to represent compact objects such as neutron stars in Entangled Relativity is indeed a good approximation of the exact solutions of Entangled Relativity when the background field goes to zero, validating the choice made by \cite{arruga:2021ep} for matching their analytical and numerical compact object solutions.

Finally, let us note that the solutions given in Eqs. (\ref{eq:msolmag}-\ref{eq:Amag}) and  Eqs. (\ref{eq:transfoERmag2elec}-\ref{eq:Aelec}) generalize the solutions provided in \cite{minazzoli:2025ep}

\section{Conclusion}
\label{sec:conc}

In this paper, we provided the generalization of the Melvin-Schwarzschild solution in Einstein-Maxwell-dilaton theories to the case of spherical compact objects with scalar hair in both the magnetic and electric cases. A possible next step could be to search for new solutions involving both electric and magnetic fields at the same time, akin to the recently discovered solutions in General Relativity \cite{barrientos:2024ej,barrientos:2025pl}. However, we anticipate significant complications arising from the scalar–matter coupling. Another direction could be to initiate the procedure described in \cite{dowker:1994pr} to generate new magnetized solutions using a different seeding hairy solution, such as the Fisher–Janis–Newman–Winicour–Levi-Civita spacetime \cite{barrientos:2024ep}.

We then used these solutions to obtain the solutions in Entangled Relativity, which is particular among the theories that can be written as an Einstein-Maxwell-dilaton theory in the \textit{Einstein frame} (Eq. (\ref{eq:actionclER})) for several, albeit very distinct, reasons:
\begin{itemize}
\item At the mathematical level, the scalar degree of freedom arising from the non-linearity of the coupling between curvature and matter is defined from the ratio between $\Lm$ and $R$ in the original frame (Eq. (\ref{eq:SER}))---see Eq. (\ref{eq:vartheta})---which we dubbed the \textit{Entangled frame} in Sec. \ref{sec:ER}. Therefore, it is (na\"ively) surprising that the theory still makes sense in the limit $R \rightarrow 0$, although it has been verified already with several exact solutions \cite{minazzoli:2021ej,minazzoli:2025ep,wavasseur:2025gg}, and again with the new solutions presented in the present paper in Eqs. (\ref{eq:msolmag}-\ref{eq:Amag}) and in Eqs. (\ref{eq:transfoERmag2elec}-\ref{eq:Aelec}). But more generally, we provided in Sec. \ref{sec:discEF} a general argument that any vacuum solution of General Relativity is a limit of a solution of Entangled Relativity when $\Lm \rightarrow 0$, albeit not the only one. 
\item At the phenomenological level, it has a built-in decoupling---which was named an \textit{intrinsic decoupling} in \cite{minazzoli:2013pr} in the framework of scalar-tensor theories---which freezes the scalar degree of freedom in many situations \cite{minazzoli:2018pr,arruga:2021pr,arruga:2021ep,minazzoli:2021cq}, thereby recovering the phenomenology of General Relativity in the situations where it has been thoroughly tested.
\item At the epistemological level, it is more parsimonious than General Relativity in terms of dimensionful universal constants \cite{minazzoli:2022ar} and therefore, arguably, preferable from Occam's razor principle, while it also better aligns with Einstein's \textit{Principle of Relativity of Inertia} \cite{einstein:1918an,einstein:1918sp,hoefer:1995cf,minazzoli:2024pp}.
\item At the theoretical level, it obliviates the Planck wall,\footnote{In the sense that in Entangled Relativity, the Planck units of length and time cannot be defined due to the lack of dimensionful constants with respect to standard physics \cite{minazzoli:2022ar}.} therefore providing a \textit{qualitatively} different road to \textit{Quantum Gravity} with respect to previous attempts \cite{minazzoli:2022ar}.
\end{itemize}

\section*{Data Availability Statement}

No Data associated in the manuscript.

\begin{acknowledgments}
We sincerely thank Eric Gourgoulhon  for his invaluable guidance for optimizing our notebooks using the open-source mathematical software SageManifolds.
\end{acknowledgments}


\bibliography{Er_sMelvin}

\begin{thebibliography}{64}%
\makeatletter
\providecommand \@ifxundefined [1]{%
 \@ifx{#1\undefined}
}%
\providecommand \@ifnum [1]{%
 \ifnum #1\expandafter \@firstoftwo
 \else \expandafter \@secondoftwo
 \fi
}%
\providecommand \@ifx [1]{%
 \ifx #1\expandafter \@firstoftwo
 \else \expandafter \@secondoftwo
 \fi
}%
\providecommand \natexlab [1]{#1}%
\providecommand \enquote  [1]{``#1''}%
\providecommand \bibnamefont  [1]{#1}%
\providecommand \bibfnamefont [1]{#1}%
\providecommand \citenamefont [1]{#1}%
\providecommand \href@noop [0]{\@secondoftwo}%
\providecommand \href [0]{\begingroup \@sanitize@url \@href}%
\providecommand \@href[1]{\@@startlink{#1}\@@href}%
\providecommand \@@href[1]{\endgroup#1\@@endlink}%
\providecommand \@sanitize@url [0]{\catcode `\\12\catcode `\$12\catcode
  `\&12\catcode `\#12\catcode `\^12\catcode `\_12\catcode `\%12\relax}%
\providecommand \@@startlink[1]{}%
\providecommand \@@endlink[0]{}%
\providecommand \url  [0]{\begingroup\@sanitize@url \@url }%
\providecommand \@url [1]{\endgroup\@href {#1}{\urlprefix }}%
\providecommand \urlprefix  [0]{URL }%
\providecommand \Eprint [0]{\href }%
\providecommand \doibase [0]{http://dx.doi.org/}%
\providecommand \selectlanguage [0]{\@gobble}%
\providecommand \bibinfo  [0]{\@secondoftwo}%
\providecommand \bibfield  [0]{\@secondoftwo}%
\providecommand \translation [1]{[#1]}%
\providecommand \BibitemOpen [0]{}%
\providecommand \bibitemStop [0]{}%
\providecommand \bibitemNoStop [0]{.\EOS\space}%
\providecommand \EOS [0]{\spacefactor3000\relax}%
\providecommand \BibitemShut  [1]{\csname bibitem#1\endcsname}%
\let\auto@bib@innerbib\@empty
\bibitem [{\citenamefont {Gibbons}\ and\ \citenamefont {ichi
  Maeda}(1988)}]{gibbons:1988nb}%
  \BibitemOpen
  \bibfield  {author} {\bibinfo {author} {\bibfnamefont {G.W.}\ \bibnamefont
  {Gibbons}}\ and\ \bibinfo {author} {\bibfnamefont {Kei}\ \bibnamefont {ichi
  Maeda}},\ }\bibfield  {title} {\enquote {\bibinfo {title} {Black holes and
  membranes in higher-dimensional theories with dilaton fields},}\ }\href
  {\doibase https://doi.org/10.1016/0550-3213(88)90006-5} {\bibfield  {journal}
  {\bibinfo  {journal} {Nuclear Physics B}\ }\textbf {\bibinfo {volume}
  {298}},\ \bibinfo {pages} {741--775} (\bibinfo {year} {1988})}\BibitemShut
  {NoStop}%
\bibitem [{\citenamefont {{Garfinkle}}\ \emph {et~al.}(1991)\citenamefont
  {{Garfinkle}}, \citenamefont {{Horowitz}},\ and\ \citenamefont
  {{Strominger}}}]{garfinkle:1991pr}%
  \BibitemOpen
  \bibfield  {author} {\bibinfo {author} {\bibfnamefont {David}\ \bibnamefont
  {{Garfinkle}}}, \bibinfo {author} {\bibfnamefont {Gary~T.}\ \bibnamefont
  {{Horowitz}}}, \ and\ \bibinfo {author} {\bibfnamefont {Andrew}\ \bibnamefont
  {{Strominger}}},\ }\bibfield  {title} {\enquote {\bibinfo {title} {{Charged
  black holes in string theory}},}\ }\href {\doibase 10.1103/PhysRevD.43.3140}
  {\bibfield  {journal} {\bibinfo  {journal} {\prd}\ }\textbf {\bibinfo
  {volume} {43}},\ \bibinfo {pages} {3140--3143} (\bibinfo {year}
  {1991})}\BibitemShut {NoStop}%
\bibitem [{\citenamefont {{Holzhey}}\ and\ \citenamefont
  {{Wilczek}}(1992)}]{holzhey:1992nb}%
  \BibitemOpen
  \bibfield  {author} {\bibinfo {author} {\bibfnamefont {Christoph F.~E.}\
  \bibnamefont {{Holzhey}}}\ and\ \bibinfo {author} {\bibfnamefont {Frank}\
  \bibnamefont {{Wilczek}}},\ }\bibfield  {title} {\enquote {\bibinfo {title}
  {{Black holes as elementary particles}},}\ }\href {\doibase
  10.1016/0550-3213(92)90254-9} {\bibfield  {journal} {\bibinfo  {journal}
  {Nuclear Physics B}\ }\textbf {\bibinfo {volume} {380}},\ \bibinfo {pages}
  {447--477} (\bibinfo {year} {1992})},\ \Eprint
  {http://arxiv.org/abs/hep-th/9202014} {arXiv:hep-th/9202014 [hep-th]}
  \BibitemShut {NoStop}%
\bibitem [{\citenamefont {Douglas}\ and\ \citenamefont
  {Kachru}(2007)}]{douglas:2007rm}%
  \BibitemOpen
  \bibfield  {author} {\bibinfo {author} {\bibfnamefont {Michael~R.}\
  \bibnamefont {Douglas}}\ and\ \bibinfo {author} {\bibfnamefont {Shamit}\
  \bibnamefont {Kachru}},\ }\bibfield  {title} {\enquote {\bibinfo {title}
  {Flux compactification},}\ }\href {\doibase 10.1103/RevModPhys.79.733}
  {\bibfield  {journal} {\bibinfo  {journal} {Rev. Mod. Phys.}\ }\textbf
  {\bibinfo {volume} {79}},\ \bibinfo {pages} {733--796} (\bibinfo {year}
  {2007})}\BibitemShut {NoStop}%
\bibitem [{\citenamefont {{Tong}}(2009)}]{tong:2009ar}%
  \BibitemOpen
  \bibfield  {author} {\bibinfo {author} {\bibfnamefont {David}\ \bibnamefont
  {{Tong}}},\ }\bibfield  {title} {\enquote {\bibinfo {title} {{Lectures on
  String Theory}},}\ }\href {\doibase 10.48550/arXiv.0908.0333} {\bibfield
  {journal} {\bibinfo  {journal} {arXiv e-prints}\ ,\ \bibinfo {eid}
  {arXiv:0908.0333}} (\bibinfo {year} {2009})},\ \Eprint
  {http://arxiv.org/abs/0908.0333} {arXiv:0908.0333 [hep-th]} \BibitemShut
  {NoStop}%
\bibitem [{\citenamefont {{Van Riet}}\ and\ \citenamefont
  {{Zoccarato}}(2024)}]{vanriet:2024pr}%
  \BibitemOpen
  \bibfield  {author} {\bibinfo {author} {\bibfnamefont {Thomas}\ \bibnamefont
  {{Van Riet}}}\ and\ \bibinfo {author} {\bibfnamefont {Gianluca}\ \bibnamefont
  {{Zoccarato}}},\ }\bibfield  {title} {\enquote {\bibinfo {title} {{Beginners
  lectures on flux compactifications and related Swampland topics}},}\ }\href
  {\doibase 10.1016/j.physrep.2023.11.003} {\bibfield  {journal} {\bibinfo
  {journal} {\physrep}\ }\textbf {\bibinfo {volume} {1049}},\ \bibinfo {pages}
  {1--51} (\bibinfo {year} {2024})},\ \Eprint {http://arxiv.org/abs/2305.01722}
  {arXiv:2305.01722 [hep-th]} \BibitemShut {NoStop}%
\bibitem [{\citenamefont {{Burgess}}\ \emph {et~al.}(2020)\citenamefont
  {{Burgess}}, \citenamefont {{Cicoli}}, \citenamefont {{Ciupke}},
  \citenamefont {{Krippendorf}},\ and\ \citenamefont
  {{Quevedo}}}]{burgess:2020fp}%
  \BibitemOpen
  \bibfield  {author} {\bibinfo {author} {\bibfnamefont {C.~P.}\ \bibnamefont
  {{Burgess}}}, \bibinfo {author} {\bibfnamefont {M.}~\bibnamefont {{Cicoli}}},
  \bibinfo {author} {\bibfnamefont {D.}~\bibnamefont {{Ciupke}}}, \bibinfo
  {author} {\bibfnamefont {S.}~\bibnamefont {{Krippendorf}}}, \ and\ \bibinfo
  {author} {\bibfnamefont {F.}~\bibnamefont {{Quevedo}}},\ }\bibfield  {title}
  {\enquote {\bibinfo {title} {{UV Shadows in EFTs: Accidental Symmetries,
  Robustness and No-Scale Supergravity}},}\ }\href {\doibase
  10.1002/prop.202000076} {\bibfield  {journal} {\bibinfo  {journal}
  {Fortschritte der Physik}\ }\textbf {\bibinfo {volume} {68}},\ \bibinfo {eid}
  {2000076} (\bibinfo {year} {2020})},\ \Eprint
  {http://arxiv.org/abs/2006.06694} {arXiv:2006.06694 [hep-th]} \BibitemShut
  {NoStop}%
\bibitem [{\citenamefont {{Hees}}\ \emph {et~al.}(2018)\citenamefont {{Hees}},
  \citenamefont {{Minazzoli}}, \citenamefont {{Savalle}}, \citenamefont
  {{Stadnik}},\ and\ \citenamefont {{Wolf}}}]{hees:2018pr}%
  \BibitemOpen
  \bibfield  {author} {\bibinfo {author} {\bibfnamefont {Aur{\'e}lien}\
  \bibnamefont {{Hees}}}, \bibinfo {author} {\bibfnamefont {Olivier}\
  \bibnamefont {{Minazzoli}}}, \bibinfo {author} {\bibfnamefont {Etienne}\
  \bibnamefont {{Savalle}}}, \bibinfo {author} {\bibfnamefont {Yevgeny~V.}\
  \bibnamefont {{Stadnik}}}, \ and\ \bibinfo {author} {\bibfnamefont {Peter}\
  \bibnamefont {{Wolf}}},\ }\bibfield  {title} {\enquote {\bibinfo {title}
  {{Violation of the equivalence principle from light scalar dark matter}},}\
  }\href {\doibase 10.1103/PhysRevD.98.064051} {\bibfield  {journal} {\bibinfo
  {journal} {\prd}\ }\textbf {\bibinfo {volume} {98}},\ \bibinfo {eid} {064051}
  (\bibinfo {year} {2018})},\ \Eprint {http://arxiv.org/abs/1807.04512}
  {arXiv:1807.04512 [gr-qc]} \BibitemShut {NoStop}%
\bibitem [{\citenamefont {{Minazzoli}}\ and\ \citenamefont
  {{Santos}}(2021)}]{minazzoli:2021ej}%
  \BibitemOpen
  \bibfield  {author} {\bibinfo {author} {\bibfnamefont {Olivier}\ \bibnamefont
  {{Minazzoli}}}\ and\ \bibinfo {author} {\bibfnamefont {Edison}\ \bibnamefont
  {{Santos}}},\ }\bibfield  {title} {\enquote {\bibinfo {title} {{Charged black
  hole and radiating solutions in entangled relativity}},}\ }\href {\doibase
  10.1140/epjc/s10052-021-09441-w} {\bibfield  {journal} {\bibinfo  {journal}
  {European Physical Journal C}\ }\textbf {\bibinfo {volume} {81}},\ \bibinfo
  {eid} {640} (\bibinfo {year} {2021})},\ \Eprint
  {http://arxiv.org/abs/2102.10541} {arXiv:2102.10541 [gr-qc]} \BibitemShut
  {NoStop}%
\bibitem [{\citenamefont {{Arruga}}\ and\ \citenamefont
  {{Minazzoli}}(2021)}]{arruga:2021ep}%
  \BibitemOpen
  \bibfield  {author} {\bibinfo {author} {\bibfnamefont {Denis}\ \bibnamefont
  {{Arruga}}}\ and\ \bibinfo {author} {\bibfnamefont {Olivier}\ \bibnamefont
  {{Minazzoli}}},\ }\bibfield  {title} {\enquote {\bibinfo {title} {{Analytical
  external spherical solutions in entangled relativity}},}\ }\href {\doibase
  10.1140/epjc/s10052-021-09818-x} {\bibfield  {journal} {\bibinfo  {journal}
  {European Physical Journal C}\ }\textbf {\bibinfo {volume} {81}},\ \bibinfo
  {eid} {1027} (\bibinfo {year} {2021})},\ \Eprint
  {http://arxiv.org/abs/2106.03426} {arXiv:2106.03426 [gr-qc]} \BibitemShut
  {NoStop}%
\bibitem [{\citenamefont {{Wambsganss}}(1998)}]{wambsganss:1998lr}%
  \BibitemOpen
  \bibfield  {author} {\bibinfo {author} {\bibfnamefont {Joachim}\ \bibnamefont
  {{Wambsganss}}},\ }\bibfield  {title} {\enquote {\bibinfo {title}
  {{Gravitational Lensing in Astronomy}},}\ }\href {\doibase
  10.12942/lrr-1998-12} {\bibfield  {journal} {\bibinfo  {journal} {Living
  Reviews in Relativity}\ }\textbf {\bibinfo {volume} {1}},\ \bibinfo {eid}
  {12} (\bibinfo {year} {1998})},\ \Eprint
  {http://arxiv.org/abs/astro-ph/9812021} {arXiv:astro-ph/9812021 [astro-ph]}
  \BibitemShut {NoStop}%
\bibitem [{\citenamefont {{Uzan}}(2011)}]{uzan:2011lr}%
  \BibitemOpen
  \bibfield  {author} {\bibinfo {author} {\bibfnamefont {Jean-Philippe}\
  \bibnamefont {{Uzan}}},\ }\bibfield  {title} {\enquote {\bibinfo {title}
  {{Varying Constants, Gravitation and Cosmology}},}\ }\href {\doibase
  10.12942/lrr-2011-2} {\bibfield  {journal} {\bibinfo  {journal} {Living
  Reviews in Relativity}\ }\textbf {\bibinfo {volume} {14}},\ \bibinfo {eid}
  {2} (\bibinfo {year} {2011})},\ \Eprint {http://arxiv.org/abs/1009.5514}
  {arXiv:1009.5514 [astro-ph.CO]} \BibitemShut {NoStop}%
\bibitem [{\citenamefont {{Will}}(2014)}]{will:2014lr}%
  \BibitemOpen
  \bibfield  {author} {\bibinfo {author} {\bibfnamefont {Clifford~M.}\
  \bibnamefont {{Will}}},\ }\bibfield  {title} {\enquote {\bibinfo {title}
  {{The Confrontation between General Relativity and Experiment}},}\ }\href
  {\doibase 10.12942/lrr-2014-4} {\bibfield  {journal} {\bibinfo  {journal}
  {Living Reviews in Relativity}\ }\textbf {\bibinfo {volume} {17}},\ \bibinfo
  {eid} {4} (\bibinfo {year} {2014})},\ \Eprint
  {http://arxiv.org/abs/1403.7377} {arXiv:1403.7377 [gr-qc]} \BibitemShut
  {NoStop}%
\bibitem [{\citenamefont {{Burns}}(2020)}]{burns:2020lr}%
  \BibitemOpen
  \bibfield  {author} {\bibinfo {author} {\bibfnamefont {Eric}\ \bibnamefont
  {{Burns}}},\ }\bibfield  {title} {\enquote {\bibinfo {title} {{Neutron star
  mergers and how to study them}},}\ }\href {\doibase
  10.1007/s41114-020-00028-7} {\bibfield  {journal} {\bibinfo  {journal}
  {Living Reviews in Relativity}\ }\textbf {\bibinfo {volume} {23}},\ \bibinfo
  {eid} {4} (\bibinfo {year} {2020})},\ \Eprint
  {http://arxiv.org/abs/1909.06085} {arXiv:1909.06085 [astro-ph.HE]}
  \BibitemShut {NoStop}%
\bibitem [{\citenamefont {collaboration}(2020)}]{abbott:2020lr}%
  \BibitemOpen
  \bibfield  {author} {\bibinfo {author} {\bibfnamefont {LVK}\ \bibnamefont
  {collaboration}},\ }\bibfield  {title} {\enquote {\bibinfo {title}
  {{Prospects for observing and localizing gravitational-wave transients with
  Advanced LIGO, Advanced Virgo and KAGRA}},}\ }\href {\doibase
  10.1007/s41114-020-00026-9} {\bibfield  {journal} {\bibinfo  {journal}
  {Living Reviews in Relativity}\ }\textbf {\bibinfo {volume} {23}},\ \bibinfo
  {eid} {3} (\bibinfo {year} {2020})}\BibitemShut {NoStop}%
\bibitem [{\citenamefont {{Fienga}}\ and\ \citenamefont
  {{Minazzoli}}(2024)}]{fienga:2024lr}%
  \BibitemOpen
  \bibfield  {author} {\bibinfo {author} {\bibfnamefont {Agn{\`e}s}\
  \bibnamefont {{Fienga}}}\ and\ \bibinfo {author} {\bibfnamefont {Olivier}\
  \bibnamefont {{Minazzoli}}},\ }\bibfield  {title} {\enquote {\bibinfo {title}
  {{Testing theories of gravity with planetary ephemerides}},}\ }\href
  {\doibase 10.1007/s41114-023-00047-0} {\bibfield  {journal} {\bibinfo
  {journal} {Living Reviews in Relativity}\ }\textbf {\bibinfo {volume} {27}},\
  \bibinfo {eid} {1} (\bibinfo {year} {2024})},\ \Eprint
  {http://arxiv.org/abs/2303.01821} {arXiv:2303.01821 [gr-qc]} \BibitemShut
  {NoStop}%
\bibitem [{\citenamefont {Penrose}(1965)}]{penrose:1965pl}%
  \BibitemOpen
  \bibfield  {author} {\bibinfo {author} {\bibfnamefont {Roger}\ \bibnamefont
  {Penrose}},\ }\bibfield  {title} {\enquote {\bibinfo {title} {Gravitational
  collapse and space-time singularities},}\ }\href {\doibase
  10.1103/PhysRevLett.14.57} {\bibfield  {journal} {\bibinfo  {journal} {Phys.
  Rev. Lett.}\ }\textbf {\bibinfo {volume} {14}},\ \bibinfo {pages} {57--59}
  (\bibinfo {year} {1965})}\BibitemShut {NoStop}%
\bibitem [{\citenamefont {Senovilla}(2012)}]{senovilla:2012bk}%
  \BibitemOpen
  \bibfield  {author} {\bibinfo {author} {\bibfnamefont {Jos{\'e} M.~M.}\
  \bibnamefont {Senovilla}},\ }\enquote {\bibinfo {title} {Singularity theorems
  in general relativity: Achievements and open questions},}\ in\ \href
  {\doibase 10.1007/978-0-8176-4940-1_15} {\emph {\bibinfo {booktitle}
  {Einstein and the Changing Worldviews of Physics}}},\ \bibinfo {editor}
  {edited by\ \bibinfo {editor} {\bibfnamefont {Christoph}\ \bibnamefont
  {Lehner}}, \bibinfo {editor} {\bibfnamefont {J{\"u}rgen}\ \bibnamefont
  {Renn}}, \ and\ \bibinfo {editor} {\bibfnamefont {Matthias}\ \bibnamefont
  {Schemmel}}}\ (\bibinfo  {publisher} {Birkh{\"a}user Boston},\ \bibinfo
  {address} {Boston},\ \bibinfo {year} {2012})\ pp.\ \bibinfo {pages}
  {305--316}\BibitemShut {NoStop}%
\bibitem [{\citenamefont {Senovilla}(2022)}]{senovilla:2022pt}%
  \BibitemOpen
  \bibfield  {author} {\bibinfo {author} {\bibfnamefont {José M.~M.}\
  \bibnamefont {Senovilla}},\ }\bibfield  {title} {\enquote {\bibinfo {title}
  {A critical appraisal of the singularity theorems},}\ }\href {\doibase
  10.1098/rsta.2021.0174} {\bibfield  {journal} {\bibinfo  {journal}
  {Philosophical Transactions of the Royal Society A: Mathematical, Physical
  and Engineering Sciences}\ }\textbf {\bibinfo {volume} {380}} (\bibinfo
  {year} {2022}),\ 10.1098/rsta.2021.0174}\BibitemShut {NoStop}%
\bibitem [{\citenamefont {Joshi}(2014)}]{joshi:2014bk}%
  \BibitemOpen
  \bibfield  {author} {\bibinfo {author} {\bibfnamefont {Pankaj~S.}\
  \bibnamefont {Joshi}},\ }\enquote {\bibinfo {title} {Spacetime
  singularities},}\ in\ \href {\doibase 10.1007/978-3-642-41992-8_20} {\emph
  {\bibinfo {booktitle} {Springer Handbook of Spacetime}}},\ \bibinfo {editor}
  {edited by\ \bibinfo {editor} {\bibfnamefont {Abhay}\ \bibnamefont
  {Ashtekar}}\ and\ \bibinfo {editor} {\bibfnamefont {Vesselin}\ \bibnamefont
  {Petkov}}}\ (\bibinfo  {publisher} {Springer Berlin Heidelberg},\ \bibinfo
  {address} {Berlin, Heidelberg},\ \bibinfo {year} {2014})\ pp.\ \bibinfo
  {pages} {409--436}\BibitemShut {NoStop}%
\bibitem [{\citenamefont {{Ludwig}}\ \emph {et~al.}(2015)\citenamefont
  {{Ludwig}}, \citenamefont {{Minazzoli}},\ and\ \citenamefont
  {{Capozziello}}}]{ludwig:2015pl}%
  \BibitemOpen
  \bibfield  {author} {\bibinfo {author} {\bibfnamefont {Hendrik}\ \bibnamefont
  {{Ludwig}}}, \bibinfo {author} {\bibfnamefont {Olivier}\ \bibnamefont
  {{Minazzoli}}}, \ and\ \bibinfo {author} {\bibfnamefont {Salvatore}\
  \bibnamefont {{Capozziello}}},\ }\bibfield  {title} {\enquote {\bibinfo
  {title} {{Merging matter and geometry in the same Lagrangian}},}\ }\href
  {\doibase 10.1016/j.physletb.2015.11.023} {\bibfield  {journal} {\bibinfo
  {journal} {Physics Letters B}\ }\textbf {\bibinfo {volume} {751}},\ \bibinfo
  {pages} {576--578} (\bibinfo {year} {2015})},\ \Eprint
  {http://arxiv.org/abs/1506.03278} {arXiv:1506.03278 [gr-qc]} \BibitemShut
  {NoStop}%
\bibitem [{\citenamefont {{Minazzoli}}(2018)}]{minazzoli:2018pr}%
  \BibitemOpen
  \bibfield  {author} {\bibinfo {author} {\bibfnamefont {Olivier}\ \bibnamefont
  {{Minazzoli}}},\ }\bibfield  {title} {\enquote {\bibinfo {title} {{Rethinking
  the link between matter and geometry}},}\ }\href {\doibase
  10.1103/PhysRevD.98.124020} {\bibfield  {journal} {\bibinfo  {journal}
  {\prd}\ }\textbf {\bibinfo {volume} {98}},\ \bibinfo {eid} {124020} (\bibinfo
  {year} {2018})},\ \Eprint {http://arxiv.org/abs/1811.05845} {arXiv:1811.05845
  [gr-qc]} \BibitemShut {NoStop}%
\bibitem [{\citenamefont {{Arruga}}\ \emph {et~al.}(2021)\citenamefont
  {{Arruga}}, \citenamefont {{Rousselle}},\ and\ \citenamefont
  {{Minazzoli}}}]{arruga:2021pr}%
  \BibitemOpen
  \bibfield  {author} {\bibinfo {author} {\bibfnamefont {Denis}\ \bibnamefont
  {{Arruga}}}, \bibinfo {author} {\bibfnamefont {Olivier}\ \bibnamefont
  {{Rousselle}}}, \ and\ \bibinfo {author} {\bibfnamefont {Olivier}\
  \bibnamefont {{Minazzoli}}},\ }\bibfield  {title} {\enquote {\bibinfo {title}
  {{Compact objects in entangled relativity}},}\ }\href {\doibase
  10.1103/PhysRevD.103.024034} {\bibfield  {journal} {\bibinfo  {journal}
  {\prd}\ }\textbf {\bibinfo {volume} {103}},\ \bibinfo {eid} {024034}
  (\bibinfo {year} {2021})},\ \Eprint {http://arxiv.org/abs/2011.14629}
  {arXiv:2011.14629 [gr-qc]} \BibitemShut {NoStop}%
\bibitem [{\citenamefont {Minazzoli}(2021)}]{minazzoli:2021cq}%
  \BibitemOpen
  \bibfield  {author} {\bibinfo {author} {\bibfnamefont {Olivier}\ \bibnamefont
  {Minazzoli}},\ }\bibfield  {title} {\enquote {\bibinfo {title} {De sitter
  space-times in entangled relativity},}\ }\href {\doibase
  10.1088/1361-6382/ac0589} {\bibfield  {journal} {\bibinfo  {journal}
  {Classical and Quantum Gravity}\ }\textbf {\bibinfo {volume} {38}},\ \bibinfo
  {pages} {137003} (\bibinfo {year} {2021})}\BibitemShut {NoStop}%
\bibitem [{\citenamefont {Wavasseur}\ \emph {et~al.}(2025)\citenamefont
  {Wavasseur}, \citenamefont {Abrial},\ and\ \citenamefont
  {Minazzoli}}]{wavasseur:2025gg}%
  \BibitemOpen
  \bibfield  {author} {\bibinfo {author} {\bibfnamefont {Maxime}\ \bibnamefont
  {Wavasseur}}, \bibinfo {author} {\bibfnamefont {Théo}\ \bibnamefont
  {Abrial}}, \ and\ \bibinfo {author} {\bibfnamefont {Olivier}\ \bibnamefont
  {Minazzoli}},\ }\bibfield  {title} {\enquote {\bibinfo {title} {Slowly
  rotating and charged black-holes in entangled relativity},}\ }\href {\doibase
  10.1007/s10714-025-03366-5} {\bibfield  {journal} {\bibinfo  {journal}
  {General Relativity and Gravitation}\ }\textbf {\bibinfo {volume} {57}}
  (\bibinfo {year} {2025}),\ 10.1007/s10714-025-03366-5},\ \Eprint
  {http://arxiv.org/abs/2411.09327} {arXiv:2411.09327 [gr-qc]} \BibitemShut
  {NoStop}%
\bibitem [{\citenamefont {Minazzoli}\ and\ \citenamefont
  {Wavasseur}(2025)}]{minazzoli:2025ep}%
  \BibitemOpen
  \bibfield  {author} {\bibinfo {author} {\bibfnamefont {Olivier}\ \bibnamefont
  {Minazzoli}}\ and\ \bibinfo {author} {\bibfnamefont {Maxime}\ \bibnamefont
  {Wavasseur}},\ }\bibfield  {title} {\enquote {\bibinfo {title} {Schwarzschild
  black-hole immersed in an electric or magnetic background in entangled
  relativity},}\ }\href {\doibase 10.1140/epjp/s13360-025-06129-y} {\bibfield
  {journal} {\bibinfo  {journal} {The European Physical Journal Plus}\ }\textbf
  {\bibinfo {volume} {140}} (\bibinfo {year} {2025}),\
  10.1140/epjp/s13360-025-06129-y}\BibitemShut {NoStop}%
\bibitem [{\citenamefont {Chehab}\ \emph {et~al.}(2025)\citenamefont {Chehab},
  \citenamefont {Minazzoli},\ and\ \citenamefont {Hees}}]{chehab:2025hl}%
  \BibitemOpen
  \bibfield  {author} {\bibinfo {author} {\bibfnamefont {Thomas}\ \bibnamefont
  {Chehab}}, \bibinfo {author} {\bibfnamefont {Olivier}\ \bibnamefont
  {Minazzoli}}, \ and\ \bibinfo {author} {\bibfnamefont {Aurelien}\
  \bibnamefont {Hees}},\ }\href {https://hal.science/hal-04996227} {\enquote
  {\bibinfo {title} {{Variation of Planck's quantum of action in the sky}},}\ }
  (\bibinfo {year} {2025}),\ \bibinfo {note} {preprint
  hal-04996227}\BibitemShut {NoStop}%
\bibitem [{\citenamefont {{Minazzoli}}(2022)}]{minazzoli:2022ar}%
  \BibitemOpen
  \bibfield  {author} {\bibinfo {author} {\bibfnamefont {Olivier}\ \bibnamefont
  {{Minazzoli}}},\ }\bibfield  {title} {\enquote {\bibinfo {title} {{Quantum of
  action in entangled relativity}},}\ }\href {\doibase
  10.48550/arXiv.2206.03824} {\bibfield  {journal} {\bibinfo  {journal} {arXiv
  e-prints}\ ,\ \bibinfo {eid} {arXiv:2206.03824}} (\bibinfo {year} {2022})},\
  \Eprint {http://arxiv.org/abs/2206.03824} {arXiv:2206.03824 [gr-qc]}
  \BibitemShut {NoStop}%
\bibitem [{\citenamefont {{Minazzoli}}(2024{\natexlab{a}})}]{minazzoli:2024mo}%
  \BibitemOpen
  \bibfield  {author} {\bibinfo {author} {\bibfnamefont {Olivier}\ \bibnamefont
  {{Minazzoli}}},\ }\bibfield  {title} {\enquote {\bibinfo {title} {{Cosmology
  in Entangled Relativity}},}\ }\href {\doibase 10.48550/arXiv.2405.02051}
  {\bibfield  {journal} {\bibinfo  {journal} {Contribution to the 2024
  Gravitation session of the 58th Rencontres de Moriond}\ ,\ \bibinfo {eid}
  {arXiv:2405.02051}} (\bibinfo {year} {2024}{\natexlab{a}})},\ \Eprint
  {http://arxiv.org/abs/2405.02051} {arXiv:2405.02051 [gr-qc]} \BibitemShut
  {NoStop}%
\bibitem [{\citenamefont {{Verde}}\ \emph {et~al.}(2024)\citenamefont
  {{Verde}}, \citenamefont {{Sch{\"o}neberg}},\ and\ \citenamefont
  {{Gil-Mar{\'\i}n}}}]{verde:2024an}%
  \BibitemOpen
  \bibfield  {author} {\bibinfo {author} {\bibfnamefont {Licia}\ \bibnamefont
  {{Verde}}}, \bibinfo {author} {\bibfnamefont {Nils}\ \bibnamefont
  {{Sch{\"o}neberg}}}, \ and\ \bibinfo {author} {\bibfnamefont {H{\'e}ctor}\
  \bibnamefont {{Gil-Mar{\'\i}n}}},\ }\bibfield  {title} {\enquote {\bibinfo
  {title} {{A Tale of Many H $_{0}$}},}\ }\href {\doibase
  10.1146/annurev-astro-052622-033813} {\bibfield  {journal} {\bibinfo
  {journal} {\araa}\ }\textbf {\bibinfo {volume} {62}},\ \bibinfo {pages}
  {287--331} (\bibinfo {year} {2024})},\ \Eprint
  {http://arxiv.org/abs/2311.13305} {arXiv:2311.13305 [astro-ph.CO]}
  \BibitemShut {NoStop}%
\bibitem [{\citenamefont {Collaboration}(2025)}]{desi:2025ar}%
  \BibitemOpen
  \bibfield  {author} {\bibinfo {author} {\bibfnamefont {DESI}\ \bibnamefont
  {Collaboration}},\ }\href {https://arxiv.org/abs/2503.14738} {\enquote
  {\bibinfo {title} {Desi dr2 results ii: Measurements of baryon acoustic
  oscillations and cosmological constraints},}\ } (\bibinfo {year} {2025}),\
  \Eprint {http://arxiv.org/abs/2503.14738} {arXiv:2503.14738 [astro-ph.CO]}
  \BibitemShut {NoStop}%
\bibitem [{\citenamefont {{Peebles}}(2025)}]{peebles:2025pt}%
  \BibitemOpen
  \bibfield  {author} {\bibinfo {author} {\bibfnamefont {Phillip James~E.}\
  \bibnamefont {{Peebles}}},\ }\bibfield  {title} {\enquote {\bibinfo {title}
  {{Status of the {\ensuremath{\Lambda}}CDM theory: supporting evidence and
  anomalies}},}\ }\href {\doibase 10.1098/rsta.2024.0021} {\bibfield  {journal}
  {\bibinfo  {journal} {Philosophical Transactions of the Royal Society of
  London Series A}\ }\textbf {\bibinfo {volume} {383}},\ \bibinfo {eid}
  {20240021} (\bibinfo {year} {2025})},\ \Eprint
  {http://arxiv.org/abs/2405.18307} {arXiv:2405.18307 [astro-ph.CO]}
  \BibitemShut {NoStop}%
\bibitem [{\citenamefont {{Einstein}}(1917)}]{einstein:1917co}%
  \BibitemOpen
  \bibfield  {author} {\bibinfo {author} {\bibfnamefont {Albert}\ \bibnamefont
  {{Einstein}}},\ }\bibfield  {title} {\enquote {\bibinfo {title}
  {{Kosmologische Betrachtungen zur allgemeinen Relativit{\"a}tstheorie}},}\
  }\href@noop {} {\bibfield  {journal} {\bibinfo  {journal} {Sitzungsberichte
  der K{\"o}niglich Preu{\ss}ischen Akademie der Wissenschaften (Berlin}\ ,\
  \bibinfo {pages} {142--152}} (\bibinfo {year} {1917})}\BibitemShut {NoStop}%
\bibitem [{\citenamefont {{Einstein}}(1918{\natexlab{a}})}]{einstein:1918an}%
  \BibitemOpen
  \bibfield  {author} {\bibinfo {author} {\bibfnamefont {A.}~\bibnamefont
  {{Einstein}}},\ }\bibfield  {title} {\enquote {\bibinfo {title}
  {{Prinzipielles zur allgemeinen Relativit{\"a}tstheorie}},}\ }\href {\doibase
  10.1002/andp.19183600402} {\bibfield  {journal} {\bibinfo  {journal} {Annalen
  der Physik}\ }\textbf {\bibinfo {volume} {360}},\ \bibinfo {pages} {241--244}
  (\bibinfo {year} {1918}{\natexlab{a}})},\ \bibinfo {note} {translation
  available at
  \url{https://einsteinpapers.press.princeton.edu/vol7-trans/49}}\BibitemShut
  {NoStop}%
\bibitem [{\citenamefont {{Einstein}}(1918{\natexlab{b}})}]{einstein:1918sp}%
  \BibitemOpen
  \bibfield  {author} {\bibinfo {author} {\bibfnamefont {Albert}\ \bibnamefont
  {{Einstein}}},\ }\bibfield  {title} {\enquote {\bibinfo {title} {{Kritisches
  zu einer von Hrn. de Sitter gegebenen L{\"o}sung der
  Gravitationsgleichungen}},}\ }\href@noop {} {\bibfield  {journal} {\bibinfo
  {journal} {Sitzungsberichte der K{\"o}niglich Preu{\ss}ischen Akademie der
  Wissenschaften (Berlin}\ ,\ \bibinfo {pages} {270--272}} (\bibinfo {year}
  {1918}{\natexlab{b}})},\ \bibinfo {note} {translation available at
  \url{https://einsteinpapers.press.princeton.edu/vol7-trans/52}}\BibitemShut
  {NoStop}%
\bibitem [{\citenamefont {{Einstein}}(1921)}]{einstein:1921bk}%
  \BibitemOpen
  \bibfield  {author} {\bibinfo {author} {\bibfnamefont {Albert}\ \bibnamefont
  {{Einstein}}},\ }\href@noop {} {\emph {\bibinfo {title} {{The meaning of
  relativity}}}}\ (\bibinfo  {publisher} {Princeton University Press},\
  \bibinfo {address} {Princeton},\ \bibinfo {year} {1921})\BibitemShut
  {NoStop}%
\bibitem [{\citenamefont {{Pais}}(1982)}]{pais:1982bk}%
  \BibitemOpen
  \bibfield  {author} {\bibinfo {author} {\bibfnamefont {Abraham}\ \bibnamefont
  {{Pais}}},\ }\href@noop {} {\emph {\bibinfo {title} {{Subtle is the Lord. The
  science and the life of Albert Einstein}}}}\ (\bibinfo  {publisher} {Oxford
  University Press},\ \bibinfo {address} {Oxford},\ \bibinfo {year}
  {1982})\BibitemShut {NoStop}%
\bibitem [{\citenamefont {{Hoefer}}(1995)}]{hoefer:1995cf}%
  \BibitemOpen
  \bibfield  {author} {\bibinfo {author} {\bibfnamefont {C.}~\bibnamefont
  {{Hoefer}}},\ }\bibfield  {title} {\enquote {\bibinfo {title} {{Einstein's
  Formulations of Mach's Principle}},}\ }in\ \href@noop {} {\emph {\bibinfo
  {booktitle} {Mach's Principle: From Newton's Bucket to Quantum Gravity}}},\
  \bibinfo {editor} {edited by\ \bibinfo {editor} {\bibfnamefont {Julian~B.}\
  \bibnamefont {{Barbour}}}\ and\ \bibinfo {editor} {\bibfnamefont {Herbert}\
  \bibnamefont {{Pfister}}}}\ (\bibinfo  {publisher} {{Birkh\"aser}},\ \bibinfo
  {address} {Boston University},\ \bibinfo {year} {1995})\ p.~\bibinfo {pages}
  {67}\BibitemShut {NoStop}%
\bibitem [{\citenamefont {{Minazzoli}}(2024{\natexlab{b}})}]{minazzoli:2024pp}%
  \BibitemOpen
  \bibfield  {author} {\bibinfo {author} {\bibfnamefont {O.}~\bibnamefont
  {{Minazzoli}}},\ }\bibfield  {title} {\enquote {\bibinfo {title} {{On the
  Principle of Relativity of Inertia in Both General and Entangled
  Relativities}},}\ }\href {\doibase 10.1134/S1063779624701132} {\bibfield
  {journal} {\bibinfo  {journal} {Physics of Particles and Nuclei}\ }\textbf
  {\bibinfo {volume} {55}},\ \bibinfo {pages} {1488--1493} (\bibinfo {year}
  {2024}{\natexlab{b}})}\BibitemShut {NoStop}%
\bibitem [{\citenamefont {{Dowker}}\ \emph {et~al.}(1994)\citenamefont
  {{Dowker}}, \citenamefont {{Gauntlett}}, \citenamefont {{Kastor}},\ and\
  \citenamefont {{Traschen}}}]{dowker:1994pr}%
  \BibitemOpen
  \bibfield  {author} {\bibinfo {author} {\bibfnamefont {Fay}\ \bibnamefont
  {{Dowker}}}, \bibinfo {author} {\bibfnamefont {Jerome~P.}\ \bibnamefont
  {{Gauntlett}}}, \bibinfo {author} {\bibfnamefont {David~A.}\ \bibnamefont
  {{Kastor}}}, \ and\ \bibinfo {author} {\bibfnamefont {Jennie}\ \bibnamefont
  {{Traschen}}},\ }\bibfield  {title} {\enquote {\bibinfo {title} {{Pair
  creation of dilaton black holes}},}\ }\href {\doibase
  10.1103/PhysRevD.49.2909} {\bibfield  {journal} {\bibinfo  {journal} {\prd}\
  }\textbf {\bibinfo {volume} {49}},\ \bibinfo {pages} {2909--2917} (\bibinfo
  {year} {1994})},\ \Eprint {http://arxiv.org/abs/hep-th/9309075}
  {arXiv:hep-th/9309075 [hep-th]} \BibitemShut {NoStop}%
\bibitem [{\citenamefont {{Junior}}\ \emph {et~al.}(2022)\citenamefont
  {{Junior}}, \citenamefont {{Yang}}, \citenamefont {{Crispino}}, \citenamefont
  {{Cunha}},\ and\ \citenamefont {{Herdeiro}}}]{junior:2022pr}%
  \BibitemOpen
  \bibfield  {author} {\bibinfo {author} {\bibfnamefont {Haroldo C.~D.~Lima}\
  \bibnamefont {{Junior}}}, \bibinfo {author} {\bibfnamefont {Jian-Zhi}\
  \bibnamefont {{Yang}}}, \bibinfo {author} {\bibfnamefont {Lu{\'\i}s C.~B.}\
  \bibnamefont {{Crispino}}}, \bibinfo {author} {\bibfnamefont {Pedro V.~P.}\
  \bibnamefont {{Cunha}}}, \ and\ \bibinfo {author} {\bibfnamefont {Carlos
  A.~R.}\ \bibnamefont {{Herdeiro}}},\ }\bibfield  {title} {\enquote {\bibinfo
  {title} {{Einstein-Maxwell-dilaton neutral black holes in strong magnetic
  fields: Topological charge, shadows, and lensing}},}\ }\href {\doibase
  10.1103/PhysRevD.105.064070} {\bibfield  {journal} {\bibinfo  {journal}
  {\prd}\ }\textbf {\bibinfo {volume} {105}},\ \bibinfo {eid} {064070}
  (\bibinfo {year} {2022})}\BibitemShut {NoStop}%
\bibitem [{\citenamefont {{Astorino}}(2013)}]{astorino2013pr}%
  \BibitemOpen
  \bibfield  {author} {\bibinfo {author} {\bibfnamefont {Marco}\ \bibnamefont
  {{Astorino}}},\ }\bibfield  {title} {\enquote {\bibinfo {title} {{Embedding
  hairy black holes in a magnetic universe}},}\ }\href {\doibase
  10.1103/PhysRevD.87.084029} {\bibfield  {journal} {\bibinfo  {journal}
  {\prd}\ }\textbf {\bibinfo {volume} {87}},\ \bibinfo {eid} {084029} (\bibinfo
  {year} {2013})},\ \Eprint {http://arxiv.org/abs/1301.6794} {arXiv:1301.6794
  [gr-qc]} \BibitemShut {NoStop}%
\bibitem [{\citenamefont {{Griffiths}}\ and\ \citenamefont
  {{Podolsk{\'y}}}(2009)}]{griffiths:2009bk}%
  \BibitemOpen
  \bibfield  {author} {\bibinfo {author} {\bibfnamefont {Jerry~B.}\
  \bibnamefont {{Griffiths}}}\ and\ \bibinfo {author} {\bibfnamefont
  {Jir{\'\i}}\ \bibnamefont {{Podolsk{\'y}}}},\ }\href@noop {} {\emph {\bibinfo
  {title} {{Exact Space-Times in Einstein's General Relativity}}}}\ (\bibinfo
  {publisher} {Cambridge University Press},\ \bibinfo {year}
  {2009})\BibitemShut {NoStop}%
\bibitem [{\citenamefont {{Cardoso}}\ and\ \citenamefont
  {{Nat{\'a}rio}}(2024)}]{cardoso:2024ar}%
  \BibitemOpen
  \bibfield  {author} {\bibinfo {author} {\bibfnamefont {Vitor}\ \bibnamefont
  {{Cardoso}}}\ and\ \bibinfo {author} {\bibfnamefont {Jos{\'e}}\ \bibnamefont
  {{Nat{\'a}rio}}},\ }\bibfield  {title} {\enquote {\bibinfo {title} {{An exact
  solution describing a scalar counterpart to the Schwarzschild-Melvin
  Universe}},}\ }\href {\doibase 10.48550/arXiv.2410.02851} {\bibfield
  {journal} {\bibinfo  {journal} {arXiv e-prints}\ ,\ \bibinfo {eid}
  {arXiv:2410.02851}} (\bibinfo {year} {2024})},\ \Eprint
  {http://arxiv.org/abs/2410.02851} {arXiv:2410.02851 [gr-qc]} \BibitemShut
  {NoStop}%
\bibitem [{\citenamefont {{Rogatko}}(2016)}]{rogatko:2016pr}%
  \BibitemOpen
  \bibfield  {author} {\bibinfo {author} {\bibfnamefont {Marek}\ \bibnamefont
  {{Rogatko}}},\ }\bibfield  {title} {\enquote {\bibinfo {title} {{Uniqueness
  of dilaton Melvin-Schwarzschild solution}},}\ }\href {\doibase
  10.1103/PhysRevD.93.044008} {\bibfield  {journal} {\bibinfo  {journal}
  {\prd}\ }\textbf {\bibinfo {volume} {93}},\ \bibinfo {eid} {044008} (\bibinfo
  {year} {2016})},\ \Eprint {http://arxiv.org/abs/1601.06577} {arXiv:1601.06577
  [hep-th]} \BibitemShut {NoStop}%
\bibitem [{\citenamefont {{Scheel}}\ \emph {et~al.}(1995)\citenamefont
  {{Scheel}}, \citenamefont {{Shapiro}},\ and\ \citenamefont
  {{Teukolsky}}}]{scheel:1995pr}%
  \BibitemOpen
  \bibfield  {author} {\bibinfo {author} {\bibfnamefont {Mark~A.}\ \bibnamefont
  {{Scheel}}}, \bibinfo {author} {\bibfnamefont {Stuart~L.}\ \bibnamefont
  {{Shapiro}}}, \ and\ \bibinfo {author} {\bibfnamefont {Saul~A.}\ \bibnamefont
  {{Teukolsky}}},\ }\bibfield  {title} {\enquote {\bibinfo {title} {{Collapse
  to black holes in Brans-Dicke theory. II. Comparison with general
  relativity}},}\ }\href {\doibase 10.1103/PhysRevD.51.4236} {\bibfield
  {journal} {\bibinfo  {journal} {\prd}\ }\textbf {\bibinfo {volume} {51}},\
  \bibinfo {pages} {4236--4249} (\bibinfo {year} {1995})},\ \Eprint
  {http://arxiv.org/abs/gr-qc/9411026} {arXiv:gr-qc/9411026 [gr-qc]}
  \BibitemShut {NoStop}%
\bibitem [{\citenamefont {{Gerosa}}\ \emph {et~al.}(2016)\citenamefont
  {{Gerosa}}, \citenamefont {{Sperhake}},\ and\ \citenamefont
  {{Ott}}}]{gerosa:2016cq}%
  \BibitemOpen
  \bibfield  {author} {\bibinfo {author} {\bibfnamefont {Davide}\ \bibnamefont
  {{Gerosa}}}, \bibinfo {author} {\bibfnamefont {Ulrich}\ \bibnamefont
  {{Sperhake}}}, \ and\ \bibinfo {author} {\bibfnamefont {Christian~D.}\
  \bibnamefont {{Ott}}},\ }\bibfield  {title} {\enquote {\bibinfo {title}
  {{Numerical simulations of stellar collapse in scalar-tensor theories of
  gravity}},}\ }\href {\doibase 10.1088/0264-9381/33/13/135002} {\bibfield
  {journal} {\bibinfo  {journal} {Classical and Quantum Gravity}\ }\textbf
  {\bibinfo {volume} {33}},\ \bibinfo {eid} {135002} (\bibinfo {year}
  {2016})},\ \Eprint {http://arxiv.org/abs/1602.06952} {arXiv:1602.06952
  [gr-qc]} \BibitemShut {NoStop}%
\bibitem [{\citenamefont {{Damour}}\ and\ \citenamefont
  {{Esposito-Farese}}(1992)}]{damour:1992cq}%
  \BibitemOpen
  \bibfield  {author} {\bibinfo {author} {\bibfnamefont {T.}~\bibnamefont
  {{Damour}}}\ and\ \bibinfo {author} {\bibfnamefont {G.}~\bibnamefont
  {{Esposito-Farese}}},\ }\bibfield  {title} {\enquote {\bibinfo {title}
  {{Tensor-multi-scalar theories of gravitation}},}\ }\href {\doibase
  10.1088/0264-9381/9/9/015} {\bibfield  {journal} {\bibinfo  {journal}
  {Classical and Quantum Gravity}\ }\textbf {\bibinfo {volume} {9}},\ \bibinfo
  {pages} {2093--2176} (\bibinfo {year} {1992})}\BibitemShut {NoStop}%
\bibitem [{\citenamefont {{Gourgoulhon}}\ \emph {et~al.}(2015)\citenamefont
  {{Gourgoulhon}}, \citenamefont {{Bejger}},\ and\ \citenamefont
  {{Mancini}}}]{gourgoulhon:2015jc}%
  \BibitemOpen
  \bibfield  {author} {\bibinfo {author} {\bibfnamefont {Eric}\ \bibnamefont
  {{Gourgoulhon}}}, \bibinfo {author} {\bibfnamefont {Michal}\ \bibnamefont
  {{Bejger}}}, \ and\ \bibinfo {author} {\bibfnamefont {Marco}\ \bibnamefont
  {{Mancini}}},\ }\bibfield  {title} {\enquote {\bibinfo {title} {{Tensor
  calculus with open-source software: the SageManifolds project}},}\ }in\ \href
  {\doibase 10.1088/1742-6596/600/1/012002} {\emph {\bibinfo {booktitle}
  {Journal of Physics Conference Series}}},\ \bibinfo {series} {Journal of
  Physics Conference Series}, Vol.\ \bibinfo {volume} {600}\ (\bibinfo
  {publisher} {IOP},\ \bibinfo {year} {2015})\ p.\ \bibinfo {pages} {012002},\
  \Eprint {http://arxiv.org/abs/1412.4765} {arXiv:1412.4765 [gr-qc]}
  \BibitemShut {NoStop}%
\bibitem [{\citenamefont {{Wavasseur}}\ and\ \citenamefont
  {{Minazzoli}}(2024{\natexlab{a}})}]{notebook_EINSTEIN_mag}%
  \BibitemOpen
  \bibfield  {author} {\bibinfo {author} {\bibfnamefont {Maxime}\ \bibnamefont
  {{Wavasseur}}}\ and\ \bibinfo {author} {\bibfnamefont {Olivier}\ \bibnamefont
  {{Minazzoli}}},\ }\href@noop {} {\enquote {\bibinfo {title}
  {{EINSTEIN\_H\_Magnetic\_BH}},}\ }\bibinfo {howpublished}
  {\url{https://github.com/mWavasseur/ER/blob/main/Art.II\%20Hairy\%20Compact\%20Objects\%20Melvin/EINSTEIN_H_Magnetic_BH.ipynb}}
  (\bibinfo {year} {2024}{\natexlab{a}}),\ \bibinfo {note} {released:
  2024-11-27}\BibitemShut {NoStop}%
\bibitem [{\citenamefont {{Wavasseur}}\ and\ \citenamefont
  {{Minazzoli}}(2024{\natexlab{b}})}]{notebook_EINSTEIN_elec}%
  \BibitemOpen
  \bibfield  {author} {\bibinfo {author} {\bibfnamefont {Maxime}\ \bibnamefont
  {{Wavasseur}}}\ and\ \bibinfo {author} {\bibfnamefont {Olivier}\ \bibnamefont
  {{Minazzoli}}},\ }\href@noop {} {\enquote {\bibinfo {title}
  {{EINSTEIN\_H\_electric\_BH}},}\ }\bibinfo {howpublished}
  {\url{https://github.com/mWavasseur/ER/blob/main/Art.II\%20Hairy\%20Compact\%20Objects\%20Melvin/EINSTEIN_H_electric_BH.ipynb}}
  (\bibinfo {year} {2024}{\natexlab{b}}),\ \bibinfo {note} {released:
  2024-11-27}\BibitemShut {NoStop}%
\bibitem [{\citenamefont {{Yazadjiev}}(2006)}]{yazadjiev:2006pr}%
  \BibitemOpen
  \bibfield  {author} {\bibinfo {author} {\bibfnamefont {Stoytcho~S.}\
  \bibnamefont {{Yazadjiev}}},\ }\bibfield  {title} {\enquote {\bibinfo {title}
  {{Magnetized black holes and black rings in the higher dimensional dilaton
  gravity}},}\ }\href {\doibase 10.1103/PhysRevD.73.064008} {\bibfield
  {journal} {\bibinfo  {journal} {\prd}\ }\textbf {\bibinfo {volume} {73}},\
  \bibinfo {eid} {064008} (\bibinfo {year} {2006})},\ \Eprint
  {http://arxiv.org/abs/gr-qc/0511114} {arXiv:gr-qc/0511114 [gr-qc]}
  \BibitemShut {NoStop}%
\bibitem [{\citenamefont {Capozziello}\ and\ \citenamefont
  {Laurentis}(2015)}]{capozziello:2015sc}%
  \BibitemOpen
  \bibfield  {author} {\bibinfo {author} {\bibfnamefont {S.}~\bibnamefont
  {Capozziello}}\ and\ \bibinfo {author} {\bibfnamefont {M.~De}\ \bibnamefont
  {Laurentis}},\ }\bibfield  {title} {\enquote {\bibinfo {title} {{F}({R})
  theories of gravitation},}\ }\href {\doibase 10.4249/scholarpedia.31422}
  {\bibfield  {journal} {\bibinfo  {journal} {Scholarpedia}\ }\textbf {\bibinfo
  {volume} {10}},\ \bibinfo {pages} {31422} (\bibinfo {year} {2015})},\
  \bibinfo {note} {revision \#147843}\BibitemShut {NoStop}%
\bibitem [{\citenamefont {{Harko}}\ and\ \citenamefont
  {{Lobo}}(2010)}]{harko:2010ep}%
  \BibitemOpen
  \bibfield  {author} {\bibinfo {author} {\bibfnamefont {Tiberiu}\ \bibnamefont
  {{Harko}}}\ and\ \bibinfo {author} {\bibfnamefont {Francisco S.~N.}\
  \bibnamefont {{Lobo}}},\ }\bibfield  {title} {\enquote {\bibinfo {title} {{f(
  R, L $_{ m }$) gravity}},}\ }\href {\doibase 10.1140/epjc/s10052-010-1467-3}
  {\bibfield  {journal} {\bibinfo  {journal} {European Physical Journal C}\
  }\textbf {\bibinfo {volume} {70}},\ \bibinfo {pages} {373--379} (\bibinfo
  {year} {2010})},\ \Eprint {http://arxiv.org/abs/1008.4193} {arXiv:1008.4193
  [gr-qc]} \BibitemShut {NoStop}%
\bibitem [{\citenamefont {{Minazzoli}}(2024{\natexlab{c}})}]{minazzoli:2024ar}%
  \BibitemOpen
  \bibfield  {author} {\bibinfo {author} {\bibfnamefont {Olivier}\ \bibnamefont
  {{Minazzoli}}},\ }\bibfield  {title} {\enquote {\bibinfo {title} {{Cosmology
  in Entangled Relativity}},}\ }\href {\doibase 10.48550/arXiv.2405.02051}
  {\bibfield  {journal} {\bibinfo  {journal} {Contribution to the 2024
  Cosmology session of the 58th Rencontres de Moriond}\ ,\ \bibinfo {eid}
  {arXiv:2405.02051}} (\bibinfo {year} {2024}{\natexlab{c}})},\ \Eprint
  {http://arxiv.org/abs/2405.02051} {arXiv:2405.02051 [gr-qc]} \BibitemShut
  {NoStop}%
\bibitem [{\citenamefont {{Wavasseur}}\ and\ \citenamefont
  {{Minazzoli}}(2024{\natexlab{c}})}]{notebook_ER_mag}%
  \BibitemOpen
  \bibfield  {author} {\bibinfo {author} {\bibfnamefont {Maxime}\ \bibnamefont
  {{Wavasseur}}}\ and\ \bibinfo {author} {\bibfnamefont {Olivier}\ \bibnamefont
  {{Minazzoli}}},\ }\href@noop {} {\enquote {\bibinfo {title}
  {{ER\_H\_Magnetic\_BH}},}\ }\bibinfo {howpublished}
  {\url{https://github.com/mWavasseur/ER/blob/main/Art.II\%20Hairy\%20Compact\%20Objects\%20Melvin/ER_H_Magnetic_BH.ipynb}}
  (\bibinfo {year} {2024}{\natexlab{c}}),\ \bibinfo {note} {released:
  2024-11-27}\BibitemShut {NoStop}%
\bibitem [{\citenamefont {{Wavasseur}}\ and\ \citenamefont
  {{Minazzoli}}(2024{\natexlab{d}})}]{notebook_ER_elec}%
  \BibitemOpen
  \bibfield  {author} {\bibinfo {author} {\bibfnamefont {Maxime}\ \bibnamefont
  {{Wavasseur}}}\ and\ \bibinfo {author} {\bibfnamefont {Olivier}\ \bibnamefont
  {{Minazzoli}}},\ }\href@noop {} {\enquote {\bibinfo {title}
  {{ER\_H\_Electric\_BH}},}\ }\bibinfo {howpublished}
  {\url{https://github.com/mWavasseur/ER/blob/main/Art.II\%20Hairy\%20Compact\%20Objects\%20Melvin/ER_H_Electric_BH.ipynb}}
  (\bibinfo {year} {2024}{\natexlab{d}}),\ \bibinfo {note} {released:
  2024-11-27}\BibitemShut {NoStop}%
\bibitem [{\citenamefont {{Wavasseur}}\ and\ \citenamefont
  {{Minazzoli}}(2024{\natexlab{e}})}]{notebook_ER_petrov}%
  \BibitemOpen
  \bibfield  {author} {\bibinfo {author} {\bibfnamefont {Maxime}\ \bibnamefont
  {{Wavasseur}}}\ and\ \bibinfo {author} {\bibfnamefont {Olivier}\ \bibnamefont
  {{Minazzoli}}},\ }\href@noop {} {\enquote {\bibinfo {title}
  {{ER\_H\_Petrov\_BH}},}\ }\bibinfo {howpublished}
  {\url{https://github.com/mWavasseur/ER/blob/main/Art.II\%20Hairy\%20Compact\%20Objects\%20Melvin/ER_H_Petrov_Classification.ipynb}}
  (\bibinfo {year} {2024}{\natexlab{e}}),\ \bibinfo {note} {released:
  2024-11-27}\BibitemShut {NoStop}%
\bibitem [{\citenamefont {Barrientos}\ \emph
  {et~al.}(2024{\natexlab{a}})\citenamefont {Barrientos}, \citenamefont
  {Cisterna}, \citenamefont {Kolář}, \citenamefont {Müller}, \citenamefont
  {Oyarzo},\ and\ \citenamefont {Pallikaris}}]{barrientos:2024ej}%
  \BibitemOpen
  \bibfield  {author} {\bibinfo {author} {\bibfnamefont {José}\ \bibnamefont
  {Barrientos}}, \bibinfo {author} {\bibfnamefont {Adolfo}\ \bibnamefont
  {Cisterna}}, \bibinfo {author} {\bibfnamefont {Ivan}\ \bibnamefont
  {Kolář}}, \bibinfo {author} {\bibfnamefont {Keanu}\ \bibnamefont
  {Müller}}, \bibinfo {author} {\bibfnamefont {Marcelo}\ \bibnamefont
  {Oyarzo}}, \ and\ \bibinfo {author} {\bibfnamefont {Konstantinos}\
  \bibnamefont {Pallikaris}},\ }\bibfield  {title} {\enquote {\bibinfo {title}
  {Mixing “magnetic” and “electric” ehlers–harrison transformations:
  the electromagnetic swirling spacetime and novel type i backgrounds},}\
  }\href {\doibase 10.1140/epjc/s10052-024-13093-x} {\bibfield  {journal}
  {\bibinfo  {journal} {The European Physical Journal C}\ }\textbf {\bibinfo
  {volume} {84}} (\bibinfo {year} {2024}{\natexlab{a}}),\
  10.1140/epjc/s10052-024-13093-x}\BibitemShut {NoStop}%
\bibitem [{\citenamefont {Barrientos}\ \emph {et~al.}(2025)\citenamefont
  {Barrientos}, \citenamefont {Cisterna}, \citenamefont {Hassaine},\ and\
  \citenamefont {Pallikaris}}]{barrientos:2025pl}%
  \BibitemOpen
  \bibfield  {author} {\bibinfo {author} {\bibfnamefont {José}\ \bibnamefont
  {Barrientos}}, \bibinfo {author} {\bibfnamefont {Adolfo}\ \bibnamefont
  {Cisterna}}, \bibinfo {author} {\bibfnamefont {Mokhtar}\ \bibnamefont
  {Hassaine}}, \ and\ \bibinfo {author} {\bibfnamefont {Konstantinos}\
  \bibnamefont {Pallikaris}},\ }\bibfield  {title} {\enquote {\bibinfo {title}
  {Electromagnetized black holes and swirling backgrounds in nonlinear
  electrodynamics: The modmax case},}\ }\href {\doibase
  https://doi.org/10.1016/j.physletb.2024.139214} {\bibfield  {journal}
  {\bibinfo  {journal} {Physics Letters B}\ }\textbf {\bibinfo {volume}
  {860}},\ \bibinfo {pages} {139214} (\bibinfo {year} {2025})}\BibitemShut
  {NoStop}%
\bibitem [{\citenamefont {Barrientos}\ \emph
  {et~al.}(2024{\natexlab{b}})\citenamefont {Barrientos}, \citenamefont
  {Cisterna}, \citenamefont {Hassaine},\ and\ \citenamefont
  {Oliva}}]{barrientos:2024ep}%
  \BibitemOpen
  \bibfield  {author} {\bibinfo {author} {\bibfnamefont {José}\ \bibnamefont
  {Barrientos}}, \bibinfo {author} {\bibfnamefont {Adolfo}\ \bibnamefont
  {Cisterna}}, \bibinfo {author} {\bibfnamefont {Mokhtar}\ \bibnamefont
  {Hassaine}}, \ and\ \bibinfo {author} {\bibfnamefont {Julio}\ \bibnamefont
  {Oliva}},\ }\bibfield  {title} {\enquote {\bibinfo {title} {Revisiting
  buchdahl transformations: new static and rotating black holes in vacuum,
  double copy, and hairy extensions},}\ }\href {\doibase
  10.1140/epjc/s10052-024-13383-4} {\bibfield  {journal} {\bibinfo  {journal}
  {The European Physical Journal C}\ }\textbf {\bibinfo {volume} {84}}
  (\bibinfo {year} {2024}{\natexlab{b}}),\
  10.1140/epjc/s10052-024-13383-4}\BibitemShut {NoStop}%
\bibitem [{\citenamefont {{Minazzoli}}\ and\ \citenamefont
  {{Hees}}(2013)}]{minazzoli:2013pr}%
  \BibitemOpen
  \bibfield  {author} {\bibinfo {author} {\bibfnamefont {Olivier}\ \bibnamefont
  {{Minazzoli}}}\ and\ \bibinfo {author} {\bibfnamefont {Aur{\'e}lien}\
  \bibnamefont {{Hees}}},\ }\bibfield  {title} {\enquote {\bibinfo {title}
  {{Intrinsic Solar System decoupling of a scalar-tensor theory with a
  universal coupling between the scalar field and the matter Lagrangian}},}\
  }\href {\doibase 10.1103/PhysRevD.88.041504} {\bibfield  {journal} {\bibinfo
  {journal} {\prd}\ }\textbf {\bibinfo {volume} {88}},\ \bibinfo {eid} {041504}
  (\bibinfo {year} {2013})},\ \Eprint {http://arxiv.org/abs/1308.2770}
  {arXiv:1308.2770 [gr-qc]} \BibitemShut {NoStop}%
\bibitem [{\citenamefont {Chauvineau}(2022)}]{chauvineau:2022pr}%
  \BibitemOpen
  \bibfield  {author} {\bibinfo {author} {\bibfnamefont {Bertrand}\
  \bibnamefont {Chauvineau}},\ }\bibfield  {title} {\enquote {\bibinfo {title}
  {Lensing by a fisher-janis-newman-winicour naked singularity: Observational
  issues related to the existence of caustic bending in the strongly scalarized
  case},}\ }\href {\doibase 10.1103/PhysRevD.105.024071} {\bibfield  {journal}
  {\bibinfo  {journal} {Phys. Rev. D}\ }\textbf {\bibinfo {volume} {105}},\
  \bibinfo {pages} {024071} (\bibinfo {year} {2022})}\BibitemShut {NoStop}%
\bibitem [{\citenamefont {{Wavasseur}}\ and\ \citenamefont
  {{Minazzoli}}(2024{\natexlab{f}})}]{notebook_appendix}%
  \BibitemOpen
  \bibfield  {author} {\bibinfo {author} {\bibfnamefont {Maxime}\ \bibnamefont
  {{Wavasseur}}}\ and\ \bibinfo {author} {\bibfnamefont {Olivier}\ \bibnamefont
  {{Minazzoli}}},\ }\href@noop {} {\enquote {\bibinfo {title}
  {Appendix\_just\_metric},}\ }\bibinfo {howpublished}
  {\url{https://github.com/mWavasseur/ER/blob/main/Art.II\%20Hairy\%20Compact\%20Objects\%20Melvin/Appendix_Just_Metric.ipynb}}
  (\bibinfo {year} {2024}{\natexlab{f}}),\ \bibinfo {note} {released:
  2024-11-27}\BibitemShut {NoStop}%
\end{thebibliography}%

\appendix

\section{Just-Fisher-Janis-Robinson-Winicour solution in the Einstein and Entangled frames}
\label{sec:FJRW}

In the Einstein-frame, the Just-Fisher-Janis-Robinson-Winicour solution in a specific radial coordinates reads \cite{arruga:2021ep,chauvineau:2022pr}
\bea \label{eq:JFJRW}
d\t s^2 &=& -\left(1-\frac{r_s}{r} \right)^b dt^2 + \left(1-\frac{r_s}{r} \right)^{-b} dr^2 \nonumber \\
&&+\left(1-\frac{r_s}{r} \right)^{1-b} r^2  \left[ d\theta^2 +\sin^2 \theta d\psi^2 \right],
\eea
with
\be
\varphi =  \frac{\sqrt{1-b^2}}{2} \ln \left(1 - \frac{\rs}{r}\right),
\ee
where $\rs$ is the Schwarzschild radius and $b \in~ ]0,1]$ is an arbitrary parameter that quantify the amount of scalar charge of the solution. Given that the trace of the Einstein field equation is $\t R = -\t T$, this solution must be such that
\be
\t R=2\t g^{\mu \nu}\partial_\mu \varphi \partial_\nu \varphi.
\ee
One can indeed verify that
\bea
\t R &=& \frac{1-b^2}{2} \rs^2 \left(1-\frac{r_s}{r} \right)^{b-2},\\
&=& 2\t g^{\mu \nu}\partial_\mu \varphi \partial_\nu \varphi.
\eea
After the conformal transformation defined and used in Sec. \ref{sec:ER}, $g_{\alpha \beta}=e^{4\alpha \varphi} \t g_{\alpha \beta}$ with $\alpha = 1/(2\sqrt{3})$---which is used in Entangled Relativity in order to go from the Einstein frame to the Entangled frame---the line element reads
\bea
\label{FJRW_ef}
ds^2 &=& -\left(1 - \frac{r_s}{r}\right)^{b - \sqrt{\frac{1 - b^2}{3}}}dt^2
+ \left(1 - \frac{r_s}{r}\right)^{-b - \sqrt{\frac{1 - b^2}{3}}}dr^2\nonumber\\
&& + \left(1 - \frac{r_s}{r}\right)^{1- b - \sqrt{\frac{1 - b^2}{3}}}r^2\left(d\theta^2 + \sin^2 \theta d\psi^2\right)\nonumber.\\
\eea
One verifies in the notebook \cite{notebook_appendix} that, surprisingly, this non-trivial line element is indeed such that 
\be
R=0.
\ee

The reason for this surprising property is given in Sec. \ref{sec:discEF}. Indeed, the field equations are such that the action defined from $\vartheta^2 R$ is minimal for $\vartheta \neq 0$, inducing that $R=0$ on-shell.\\

There is one class of scalar-tensor theories for which that would be true, and it comes to the fact that
\be \label{eq:trascog}
\forall~~\t g_{\alpha \beta} = \phi g_{\alpha \beta}\textrm{, with } \phi = e^{\pm 2\varphi/\sqrt{3}},
\ee
one has
\be \label{eq:trascogR}
\sqrt{-g} \phi R = \sqrt{- \t g} \left[\t R - 2 \t g^{\alpha \beta} \partial_\alpha \varphi \partial_\beta \varphi \right].
\ee
As one can check with Eq. (\ref{eq:conftransER})---see \cite{minazzoli:2021ej} for more details---Entangled Relativity belongs to this class of theories.
One can notice from both Eq. (\ref{eq:trascog}) and (\ref{eq:trascogR}) that the Lagrangian density in the Einstein frame possesses a $\mathbb{Z}_2$ symmetry ($\varphi \rightarrow -\varphi$), which corresponds to an inversion symmetry ($\phi \rightarrow \phi^{-1}$) of the conformal transformation. Hence, we can deduce another metric from a conformal transformation of the metric in Eq. (\ref{eq:JFJRW}), by using the inverse conformal transformation with respect to the one previously considered---that is, $\hat g_{\alpha \beta}=e^{-4\alpha \varphi} \t g_{\alpha \beta}$ with $\alpha = 1/(2\sqrt{3})$. The resulting line element reads
\bea
\label{FJRW_ef2}
d \hat s^2 &=& -\left(1 - \frac{r_s}{r}\right)^{b + \sqrt{\frac{1 - b^2}{3}}}dt^2
+ \left(1 - \frac{r_s}{r}\right)^{-b + \sqrt{\frac{1 - b^2}{3}}}dr^2\nonumber\\
&& + \left(1 - \frac{r_s}{r}\right)^{1- b + \sqrt{\frac{1 - b^2}{3}}}r^2\left(d\theta^2 + \sin^2 \theta d\psi^2\right)\nonumber.\\
\eea
One indeed verifies that $\hat R =0$ (see \cite{notebook_appendix}). Obviously, the Lagrangian density in the Einstein frame also possesses a shift symmetry ($\varphi \rightarrow \varphi \pm \varphi_0$, where $\varphi_0$ is a constant), which corresponds to a normalization change of the conformal transformation ($\phi \rightarrow \phi/ \phi_0 := \phi ~e^{\mp 2 \varphi_0/\sqrt{3}}$).

Let us note that both the $\mathbb{Z}_2$ and shift symmetries are broken when $\Lm \neq 0$ for theories corresponding to Eq. (\ref{eq:trascogR}) with a multiplicative scalar-matter coupling in the Lagrangian density. Therefore, in Entangled Relativity, both symmetries become approximate symmetries only when $\Lm \rightarrow 0$. 

\end{document}